\documentclass[twocolumns,8pt]{IEEEtran}

\usepackage{cite}

\usepackage{bbm}
\usepackage{siunitx}
\usepackage{epsfig}
\usepackage{textcomp}
\usepackage{xcolor}
\usepackage{amsfonts,amsmath,color,amssymb,amsxtra,graphicx}
\usepackage{amsmath}
\usepackage{amssymb}
\usepackage{bm}
\usepackage{cuted}
\usepackage{stackengine}
\usepackage{subfigure}
\usepackage{graphicx}

\usepackage{algorithm}
\usepackage{setspace}
\usepackage{algpseudocode}
\usepackage{clipboard}
\usepackage{amsthm}

\newtheorem{lemma}{Lemma}

\newtheorem{theorem}{Theorem}
\newclipboard{myclipboard}

\newcommand{\eq}[1]{(\ref{#1})}

\newcommand{\subparagraph}{}
\usepackage{titlesec}
\usepackage{lipsum}

\titlespacing{\section}{0pt}{1ex plus 1ex minus 0.2ex}{1ex plus 0.2ex}
\titlespacing{\subsection}
{0pt}{1ex plus 1ex minus .2ex}{1ex plus .2ex}

\usepackage{eqparbox}
\usepackage{caption}

\makeatletter
\def\thm@space@setup{\thm@preskip=0pt
	\thm@postskip=0pt}
\makeatother

\IEEEoverridecommandlockouts
\begin{document}

\title{Age-of-Information
with Information Source Diversity
 in an Energy Harvesting System}

\author{Elvina~Gindullina, ~Leonardo Badia ~\IEEEmembership{Senior Member,~IEEE,} and ~Deniz G\"{u}nd\"{u}z  ~\IEEEmembership{Senior Member,~IEEE}\\
\thanks{E. Gindullina and L. Badia are with the University of Padova, Dept.\ of Information Engineering, via Gradenigo 6B, 35131 Padova, Italy, email: \{gindullina, badia\}@dei.unipd.it }
\thanks{D. G\"{u}nd\"{u}z is with the Imperial College London, Dept. of Electrical and Electronic Engineering, London SW7 2AZ, U.K., email: d.gunduz@imperial.ac.uk}
\thanks{This work has received funding from the European Union's Horizon 2020 research and innovation programme under the Marie Sk\l odowska-Curie grant agreement No. 675891 \mbox{(SCAVENGE)}. D. G\"{u}nd\"{u}z also received funding from the European Research Council (ERC) through project BEACON (grant No. 677854). Part of this work was presented at \cite{gindullina2019average}.}
\thanks{Manuscript received April 23, 2020; revised October 29, 2020 and May 5, 2021; accepted June 09, 2021.}
}


\maketitle

\begin{abstract}
Age of information (AoI) is a key performance metric for the Internet of things (IoT). Timely status updates are essential for many IoT applications; however, they often suffer from harsh energy constraints and the unreliability of underlying information sources. To overcome these unpredictabilities, one can employ multiple sources that track the same process of interest, but with different energy costs and reliabilities.
We consider an energy-harvesting (EH) monitoring node equipped with a finite-size battery and collecting status updates from multiple heterogeneous information sources. We investigate the policies that minimize the average AoI, formulating a Markov decision process (MDP) to choose the optimal actions of either updating from one of the sources or remaining idle, based on the current energy level and the AoI at the monitoring node. We analyze the structure of the optimal solution for different cost/AoI distribution combinations, and compare its performance with an aggressive policy that transmits whenever possible.
\end{abstract}


\section{Introduction}

Internet of things (IoT) systems are increasingly being exploited for a variety of applications that encompass every aspect of our lives \cite{javed2018internet}. In many of these applications freshness of the monitored information can play an important role for the system performance. Age of information (AoI) is a key performance indicator in mission-critical and time-sensitive applications, including smart transportation, healthcare, remote surgery, robotics cooperation, public safety, industrial process automation, to count a few. AoI quantifies the freshness of knowledge about the status of the system being monitored \cite{kaul2012real}, \cite{kaul2011minimizing}. For instance, in autonomous driving, timely collection of traffic information and vehicle-generated data is essential for the safety of all road users. Another important example is factory automation, where real-time control of production also requires timely delivery of status updates \cite{zhang2015mission}.

One limitation against frequent updates is the energy supply of the sensor. Since sensing devices are typically wireless, and often placed in remote areas, it would be impractical to power them through cables. If the device is powered only through batteries, a significant downtime would hinder the provision of reliable and up-to-date information. In these situations, broad autonomy for reliable IoT systems can be obtained through energy harvesting (EH) combined with rechargeable batteries. This, however, would further require a smart sensing and communication strategy \cite{gunduz2014designing}. 
Indeed, the integration of energy harvesters reduces the maintenance cost of IoT and increases the energy self-sustainability, but comes at a price of not guaranteeing uninterrupted operation of the device. 

We focus on an EH monitoring node, whose goal is to track the underlying process as closely as possible, i.e., with the minimum average AoI, within the constraints of stochastic energy arrivals from ambient sources of energy and a finite battery capacity. Also, we consider the role of multiple information sources that monitor the same underlying process of interest called \textit{information source diversity}, where each source provides a different trade-off between the cost of sensing and the freshness of the provided status update. Hence, the policy governing the operation of the system does not simply make a binary choice between providing a new status update or not, but must also include the optimal choice of the specific information source to be used. To clarify, the policy might also choose to wait, instead of updating immediately in a myopic fashion, in order to accumulate energy so as to be able to use a more reliable information source in the future. 

We compare the performance of the optimal policy with a greedy ``aggressive'' update policy in terms of average AoI, highlighting the situations where optimization is really needed as opposed to the simple implementation of an ``update whenever needed'' strategy. We also quantify the additional gains in the minimal long-term average AoI due to multiple information sources, as well as how the quality of these sources affects the outcome. Finally, we compute the power expenditure of these policies, and discuss how the added dimensionality of the problem affects the system performance.

 
 
\subsection{Background}

Several papers study the average AoI minimization with a single energy-harvesting source  \cite{ceran2019reinforcement, wu2017optimal, arafa2019age, arafa2018age1,bacinoglu2019optimal,dabiri2018average,feng2021age,bacinoglu2018achieving,leng2019age,feng2018minimizing,farazi2018average,farazi2018age,arafa2018online}, whereas very few papers are focused on the average AoI with multiple information sources. In \cite{yates2018age, yates2012real},  authors consider a system where independent sources send status updates through a shared first-come-first-serve M/M/1 queue to a monitor, and find the region of feasible average status ages for two and multiple sources. Similarly, in \cite{sun2018age}, a system with $n$ sources is considered to provide status updates to multiple servers via a common queue. The authors formulate an AoI minimization problem and propose online scheduling policies. In \cite{buyukates2019age}, a single source node transmits status updates of two types to multiple receivers. The authors determine the optimal stopping thresholds  to individually and jointly optimize the average age of two-type updates at the receiver nodes. In \cite{pappas2015age}, a multi-objective formulation is proposed for scheduling transmissions in a system with multiple information sources that monitor different processes. The objective is to balance the AoI of these different processes. Similarly, in  \cite{tripathi2017age}, the AoI minimization problem is also formulated for a system with multiple information sources that monitor different processes, and a monitoring node that communicates with the information sources through orthogonal channels. The authors propose the policy that converts the scheduling problem into a bipartite matching problem between the sets of channels and sensors.  In \cite{hsu2019scheduling}, the authors study the scenario where a base station updates many network users. New information is randomly generated, and the base station can serve at most one user for each transmission. 
A structural MDP scheduling algorithm and an index scheduling algorithms were introduced.  

  \begin{figure}[!t]	
	\centering
	\includegraphics[width=0.55\columnwidth]{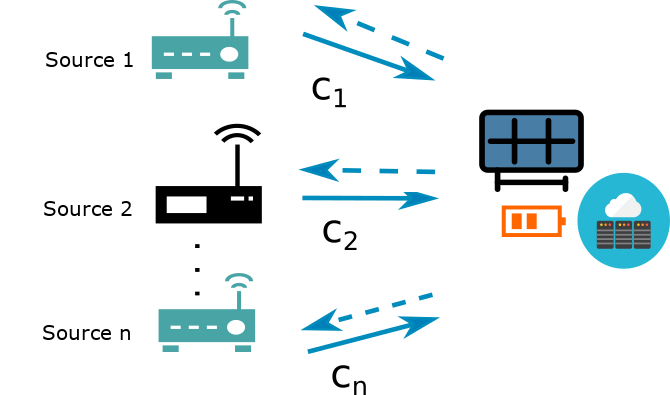}
	\caption{System model consisting of $n$ information sources.}
	\label{fig:scheme}	
	\vspace{-0.3cm}
\end{figure}

 One of the main challenges of  deriving age-optimal transmission policies using MDP-based formulation is the large size of the state space of the  system. This problem has been extensively studied for single-source systems. One of the ways to tackle this challenge is by demonstrating the optimality of  a threshold-policy.
In \cite{abd2019online}, the authors study a real-time IoT-enabled monitoring system in which a source node is responsible for maintaining the freshness of information status at a destination node. The source node is powered by wireless energy transfer. The authors adopt an MDP approach and characterize the throughput-optimal policy. In \cite{leng2019age}, the authors study the average AoI in EH cognitive radio communications, where the secondary user, i.e., EH sensor, performs spectrum sensing and status updates in a way that minimizes the average AoI based on its energy availability and the availability of the  primary spectrum. The problem is formulated as a partially observable Markov decision process, and the optimal sensing and updating policies are shown to have threshold structure. The structural properties of the optimal policy for a single IoT device, where an IoT device updates the destination node via the wireless channel, are analysed in \cite{zhou2019joint}. The authors consider a scenario where  joint status sampling and updating process is designed to minimize the average AoI at the destination.  The problem is formulated as an infinite horizon average cost constrained MDP that is transformed into an unconstrained MDP using a Lagrangian method. For the single IoT device, the optimal policy is shown to be of threshold type. Similar scenario is considered in \cite{wang2020minimizing}, where an IoT device is classified as a secondary user that exploits the spectrum opportunities of the licensed band and updates the destination node. 

Instead, the dimensionality problem in multi-source systems is tackled in  \cite{abd2020reinforcement}, where the authors  consider a multi-source RF-powered communication system and propose a reinforcement learning framework for optimizing the AoI. 

\subsection{Our Contributions}

In this work, we consider a specific kind of multi-source system, where the status updates are generated upon request by an energy-harvesting \emph{monitoring node} using multiple heterogeneous information \emph{sources} that monitor the same underlying process. These different sources may capture different physical phenomena from an abstract perspective. For example, there may be multiple sensors monitoring the same process of interest using distinct technologies for the transducers, thus resulting in different accuracies and costs. Alternatively, the
heterogeneity of the sources may stem from different channels that may convey the information (i.e., by means of different technologies,
routes, communication links, or all of the above). Thus, each of the sources offers its own tradeoff of energy vs.\ age, resulting
in information source diversity, and the monitoring node may seek to optimize the resulting AoI over time.
This ought to take into account a constrained energy budget and the characteristics of all the information sources. In our model, each source may have available updates with different ages, due to its sampling of the underlying process at possibly diverse rates.

A sample scenario is crowdsensing, in which AoI can play an important role when choosing the source of updates. In crowdsensing, a monitor and some users are connected via the cloud \cite{yang2016incentive}. The monitor sends the sensing task description to the users, and receives sensing plans, based on which to perform user selection. The AoI received from each information source depends on multiple factors such as sampling frequency, continuity of energy arrivals to the source nodes (assuming that the source nodes are powered by the ambient energy sources \cite{altinel2019modeling}), channel state, delay, and, in general, the robustness of a node.

Our first contribution is to analyze different heterogeneous information sources, and study how the combinations of cost and age distribution affect the resulting average AoI. We investigate the behavior of the optimal solution, which depends on the configuration of the information sources, through numerical analysis. In contrast to previous results in the literature \cite{abd2019online, leng2019minimizing, zhou2019joint, wang2020minimizing}, we show through examples that the optimal policy exhibits a threshold behavior only versus the AoI but in general not when the energy increases, since sometimes it may be convenient to refrain from updating and instead cumulating energy for a later update from a more expensive source. 

As another contribution, we compare the performance of the optimal and aggressive policies, and find the threshold of the EH rate in which it is reasonable to apply the aggressive policy. We evaluate the effect of an increase in the average system cost on the performance. 
Finally, we assess if an increase in the number of  information sources affects the overall performance.

\subsection{Organization of the Paper}
 
The rest of this paper is organized as follows. In Section \ref{sec:mod}, the system description, problem formulation, and solution approaches are introduced. Numerical results are presented in Section \ref{sec:res}, providing a comparison with an aggressive policy. The paper is concluded in Section \ref{sec:con}, where possible further developments are also outlined.

\section{System Model and Problem Formulation}
\label{sec:mod}

We focus on a communication system formed by a single energy-harvesting monitoring node and $n$ heterogeneous information sources, all capable of measuring the status of an underlying process. 
The monitoring node can query any of these information sources to receive an update on the status of the underlying process. For example, these information sources may model sensors with different technologies measuring the same process. In this paper, we consider such a general scenario, that could be further detailed to a multi-sensor, multi-radio, or multi-transducer scenarios \cite{himayat2014multi, qiu2018can}. Time is divided into slots of equal length, and we assume that the monitoring node can query from only one of the sources in each time slot. The received status update becomes available at the beginning of the next time slot. We highlight two important dynamics at the monitoring node: energy fluctuations and the AoI. The objective is to minimize the average AoI at the monitoring node taking into account the time-varying energy budget.


We assume that the monitoring node is equipped with a rechargeable battery of finite capacity $B$, and can harvest energy from ambient sources. Fluctuations in the battery  of the monitoring node are defined by two processes: harvested energy in each time slot and the energy consumption caused by the queries for a status update. Energy harvested over time is represented as an independent and identically distributed (i.i.d.) binary random process $\{e(t)\}_{t=1}^\infty$. At each time slot $t$ the monitoring node receives $e(t) \in \{0, \bar{e}\}$ energy units, such that $P(e(t) = \bar{e}) = \lambda$. 

The energy cost of requesting an update from source $i$, $i \in [n] \triangleq \{1,2, ..., n\}$, is denoted by $c_i$, a collective value that reflects the energy consumption of the monitoring node to acquire an update from source $i$. This may include the cost of sending a request and receiving an update if the sources are remote sensors, or simply the cost of operating that sensor if they are local.
For simplicity, we consider $c_i \in \mathbb{Z}^+$ corresponding to integer multiples of a unit of energy.


The AoI at time $t$, denoted by $\delta(t)$, refers to the age of the most recent status update available at the monitoring node 
\cite{ceran2019average}. If a more recent update is not received, $\delta(t)$ is increased by 1 at each time slot. We assume that $\delta(t) \in [0, 1, ..., \delta_{\rm max}]$, as any AoI beyond $\delta_{\rm max}$ has the same utility for the system, which reduces the dimensionality of the problem. 

The status updates provided by the information sources are not necessarily \emph{fresh}, i.e., with zero age. Due to various factors, such as the sensing technology or the processing of the measurements, we assume that the status updates may have different ages when they arrive at the monitoring node. We consider probabilistic AoI for the updates received from each information source; that is, we assume that the source nodes provide status updates with ages within the interval [$\alpha$, $\beta$] ($\alpha{<} \beta$), where $\alpha$ is the most \emph{fresh} status update while  $\beta$ is the most \emph{stale} one, typically with different distributions.  
We assume that $\alpha \geq1$, in order to incorporate the transmission time of the status update.

To model the different AoI distributions from each source, denote by  $\gamma_{i,j}$ the probability of receiving a status update of age $j$ from source $i$, where $j\in [\alpha, \beta]$ and $i \in [n]$.


It is reasonable to assume that the sources with higher probability to deliver a fresh status update have a higher energy cost. Otherwise, a source which is both more costly and provides more stale state updates would never be used, and can safely be removed from the system model.  


\subsection{Markov Decision Process (MDP) Formulation}

We aim to determine the policy that minimizes the average AoI at the monitoring node. To achieve this, the monitoring node optimally chooses the action to take at each time slot. Possible actions include requesting an update from one of the information sources at the beginning of each time slot, or staying idle. This choice is made taking into account the battery level and the age of the most recent status update available at the monitoring node. This problem can be formulated as an MDP, consisting of a tuple ${<}\mathcal{S}, \mathcal{A}, P, R{>}$, where:
\begin{itemize}
	\item $ \mathcal{S}$ is the state space where the process evolves;
	\item $ \mathcal{A}$ is the set of actions to control the state dynamics;
	\item $P$ denotes the state transition probability function;
	\item $R$ is the reward function defined on state transitions. 
\end{itemize}

The action taken by the monitoring node at time $t$ is denoted by $a(t)$, chosen from a finite action space $\mathcal{A} {=} \{a_0, a_1, a_2, ..., a_n\}$, where $a_i$ corresponds to querying source $i$ for an update,  $i {\in} [n]$, while $a_0$ corresponds to remaining idle. 
The system state is described by the pair of variables $s(t) = (b(t), \delta(t))$, $\delta(t) \in  [\delta_{\max}]$ and $b(t) \in \lfloor B \rfloor \triangleq \{0, 1, ..., B\}$. 
%
We denote by $\bar{\delta}(t)$ the age of the status update received at time $t$. Note that $\bar{\delta}(t)$ is a random variable depending on action $a(t)$. We set $\bar{\delta}(t) = \delta_{\rm max}$ if $a(t) = a_0$. Moreover, if $\bar{\delta}(t)$ happens to be larger than the age of the already available status information, $\delta(t-1)+1$, the current value is kept and no update is performed.  Thus, the AoI is updated as:
\begin{equation}
\delta(t) = \min \big\{ \delta(t-1)+1, \bar{\delta}(t), \delta_{\max} \big\}.
\label{eq:updatedelta}
\end{equation}
The energy level in the battery $b(t)$ at time $t$ evolves according to the cost of an action taken and the harvested energy within that time slot:
\begin{equation}
b(t)=\min \left\{ b(t-1) - \sum_{i=1}^n c_i \cdot \mathbbm{1}{(a(t) = a_i)} + e(t), B \right\},
\end{equation}
where $\mathbbm{1}(x)$ is the indicator function: $\mathbbm{1}(x) = 1$ when $x$ holds, and $\mathbbm{1}(x) = 0$ otherwise. Action $a_i$ is not allowed if $b(t) {<} c_i$, $i \in [n]$. We have a finite state space of dimension $\delta_{\rm max} \cdot(B+1)$.


The transition probabilities are given below for $a_i \in \{a_1, a_2, ..., a_n\}$, and $\bar{\delta(t)} \in \{\alpha, \alpha {+} 1, ..., \beta \}$.
\begin{equation}
\begin{cases} 
P\big[s(t+1) =  (\min\{b {+} \bar{e} {-} c_{i}, B\}, \min\{j, \delta{+}1, \delta_{\rm max}\}) \\
\hspace{0.2cm}| s(t) =(b, \delta), a(t) =a_i\big] = \lambda \gamma_{i,j} \hspace{0.75cm}  \text{for $b\geq c_i$, $j \in [\alpha, \beta],$}\\	
P\big[s(t+1) =(b {-} c_{i}, \min\{j, \delta{+}1, \delta_{\rm max}\}\}) \\
\hspace{0.2cm} |s(t) =(b, \delta), a(t) =a_i\big] = (1{-}\lambda) \gamma_{i,j}   \hspace{0.2cm}  \text{for $b\geq c_i$, $j \in [\alpha, \beta]$},\\
\end{cases}
\label{tr_ai}
\end{equation}

When  the node stays idle, i.e., $a(t){=}a_0$, the transition probabilities take the following form:
	\begin{equation}
\begin{array}{rcll}
	P\big[s(t+1) \!\!&\!\!{=}\!\!&\!\! (b, \min\{\delta+1, \delta_{\rm max}\}) \\
        &&   | s(t) =(b, \delta), a(t) = a_0\big] = 1 - \lambda  \;\; & \text{$b{<}B$} \\
	P\big[s(t+1) \!\!&\!\!{=}\!\!&\!\! (\min\{b + \bar{e}, B\}, \min\{\delta+1, \delta_{\rm max}\}) \\
	&&   | s(t) = (b, \delta), a(t) = a_0\big] = \lambda   \;\; &   \text{$b{<}B$} \\
	P\big[s(t+1) \!\!&\!\!{=}\!\!&\!\! (B, \min\{\delta+1, \delta_{\rm max}\}) \\
	&&   | s(t) = (B, \delta), a(t) = a_0\big] = 1  \;\; 
	\end{array}
	\label{tr_a0}
	\end{equation}
Note that when the monitoring node chooses to stay idle and its energy storage is full (i.e., $B=b$), the state transition only involves the increase in the AoI since no more energy can be stored in the battery, therefore this transition is deterministic.
The reward received at time $t$ depends on the action chosen and the age of the update received at the monitoring node:
\begin{equation}
	R(s(t+1)|s(t), a(t) = a_i) = \delta(t+1).
	\label{eq:reward}
\end{equation}

The problem is framed as a first-order Markovian dynamics as the next state depends only on the current state $s(t)$ and the current action $a(t)$. 

The deterministic stationary  policy $\pi: \mathcal{S} \rightarrow \mathcal{A}$  defines an action $a(t)$ at each time slot depending on the current state. A stationary policy $\pi$  means that $\pi_i = \pi$  for all  $t = 1, 2, ....$;  we let $\delta^{\pi}_t$ denote the sequence of AoI caused by policy $\pi$. 
The infinite-horizon time-average AoI, when policy $\pi$ is employed, starting from initial state $s_0$, is defined as \cite{ceran2019average}:
\begin{equation}
V^{\pi}(s_0) = \lim \sup_{T \rightarrow \infty} \frac{1}{T} \mathbbm{E}\left[ \sum_{t=0}^T \delta^{\pi}(t) | s(0) = s_0
\right].
\end{equation}

A policy is \emph{optimal} if it minimizes the infinite-horizon average AoI - $V^{\pi}(s)$:
\begin{equation}
V(s) = \min_{\pi} V^{\pi}(s).
\end{equation}
To solve this optimization, we can use the offline dynamic
programming approach, which is a quite common methodology successfully
used in other problems related to efficient exploitation of harvested
energy \cite{fawaz2018optimal} and can be solved via standard techniques such as
Value Iteration \cite{puterman2014markov}. In the offline approach, we model the state transition function based on the prior knowledge of the age statistics of the updates received from different sources, $\gamma_{i, j}$, and the environmental characteristics, $\lambda$. The solution represents the map of actions to be chosen in different states. 

\begin{algorithm}[!t]
	\caption{Relative VI Algorithm}\label{VI}
	\small
	\begin{algorithmic}
		\State set $v^0(s) = 0$ $\forall s \in \mathcal{S}$
		\State set n = 1, $\epsilon {>} 0$  
		\State set $V^0(s) = 0$ $\forall s \in \mathcal{S}$
		\Repeat
		\State $n \leftarrow n+1$ 
		\ForAll {$s \in \mathcal{S}$}
		\begin{equation*}
			\begin{split}
				\quad v^n(s) &= \min_{a \in \mathcal{A}} \sum_{s' \in \mathcal{S}}P(s'|s,a) \Big[\delta(s'|s,a) + V^{n-1}(s')\Big] \\
				\quad V^n(s) &= v^n(s) - v^n(s_0)
			\end{split}
		\end{equation*}
		\text{$\quad$ where $s_0$ is a fixed state chosen arbitrarily}
		\EndFor
		\Until {$sp(V^n - V^{n - 1}) {<} \epsilon$}
		\State \Return ${\rm arg} \min V(s)$
	\end{algorithmic}
\vspace{-0.1cm}
\end{algorithm}

\section{Performance evaluation}
\label{sec:res}

We compare the effect of different cost combinations and cost-reliability dependencies on the performance of different policies. We consider the cost distribution of information sources, the age distribution of updates received from different sources, and the parameter of the EH process, $\lambda$. 

To validate the optimal approach, we compare its performance with that of the aggressive policy, which requests a status update at each time slot from the most costly information source that its current battery state affords. The optimal solution is obtained via the value iteration (VI) algorithm described in \cite{puterman2014markov}, which we also provide  in Algorithm \ref{VI} for completeness. The optimal stationary deterministic policy obtained by Algorithm \ref{VI} specifies the decision rules that maps  the current energy level and AoI to deterministic actions.

In Algorithm \ref{VI}, $sp(V^n - V^{n - 1}) {<} \epsilon$ is a stopping criterion, where $sp(V) \triangleq \max_{s \in \mathcal{S}} V(s) - \min_{s \in \mathcal{S}} V(s)$. We run the relative VI algorithm until the stopping criterion holds. At that moment the policy $\pi$ achieves an average-cost AoI that is within $\epsilon \cdot 100\%$ of optimal.

\subsection{Impact of different cost functions}

Since our model and formulation are fairly general, the cost of requesting an update may result from very different reasons (sampling, processing and/or communication costs). Therefore, it is difficult to provide precise cost values and their relation across sources.  We assume that the energy cost of any source takes values between $c_{\rm min}$ and $c_{\rm max}$, where  $0<c_{\rm min} < c_{\rm max} < B$. In this way, we guarantee that all the sources  are available to the monitoring node to query a status update as long as there is sufficient energy in the battery. In particular, we set the values of $c_{\rm min} = 0.05B$, $c_{\rm max} = 0.95B$.

\begin{figure}[!t]%
	\centering
	\includegraphics[width=0.45\textwidth]{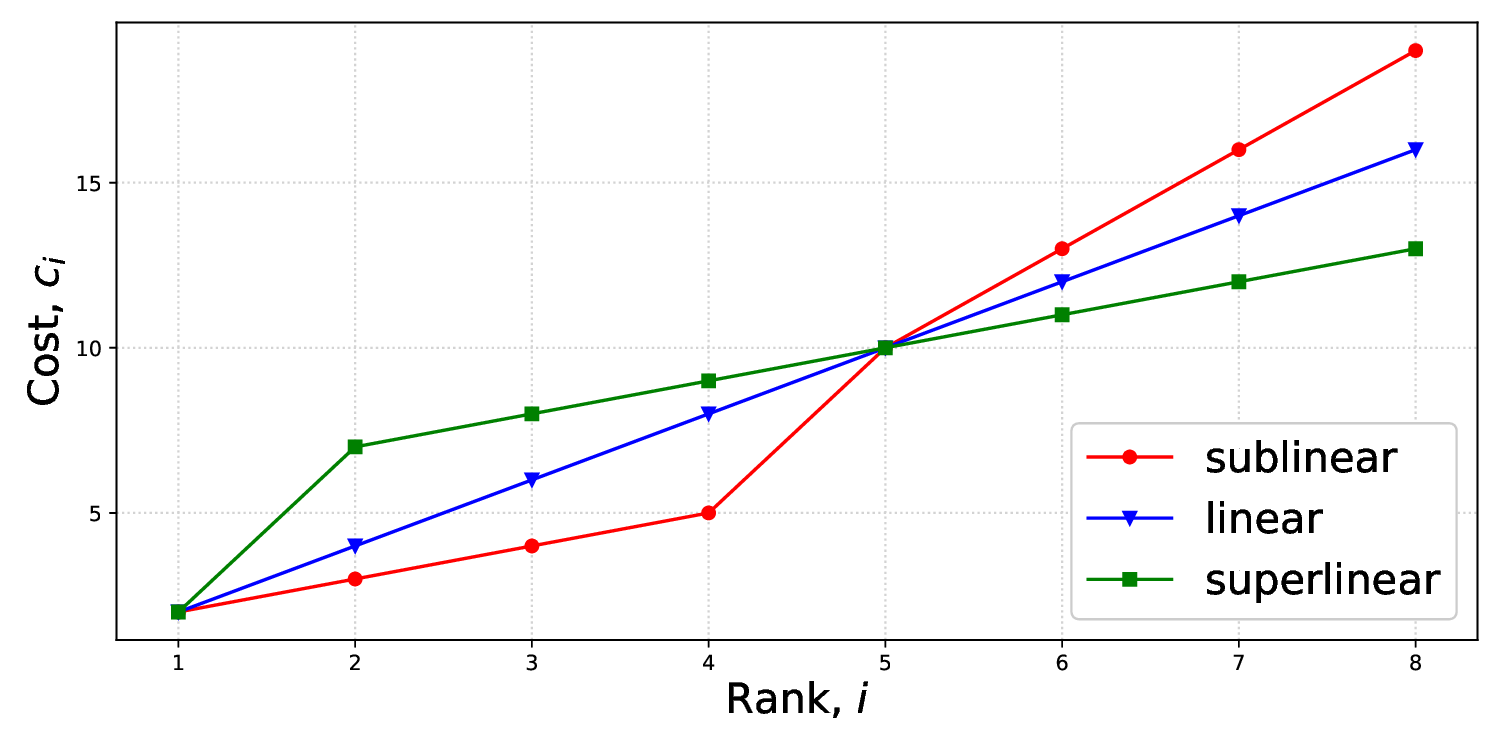}
	\vspace{-0.2cm}
	\caption{Rank-cost dependencies.}
	\label{fig:rank_cost}%
	\vspace{-0.3cm}
\end{figure}

To evaluate the effect of different cost combinations of the sources, we consider three cases, as per Fig. \ref{fig:rank_cost}, each with the same average cost value: \textit{superlinear}, \textit{linear} and \textit{sublinear}. In Fig. \ref{fig:rank_cost}, term `Rank' corresponds to the index of source $i$, such that a source with a higher index has a higher rank and higher cost, respectively. The aforementioned dependencies do not carry any special ``physical meaning'', they are simply chosen to investigate the impact of cost values on the average AoI. Indeed, other functions can also be used. Obviously, changing the average cost will affect the average AoI, but the effect of concavity on the target metric is not obvious. Thus, we focus on these trends to analyse the effect of  ``concavity" on the average AoI and also for easier reproducibility of our results.

\begin{figure}[!t]
	\centering
	\includegraphics[width=0.45\textwidth]{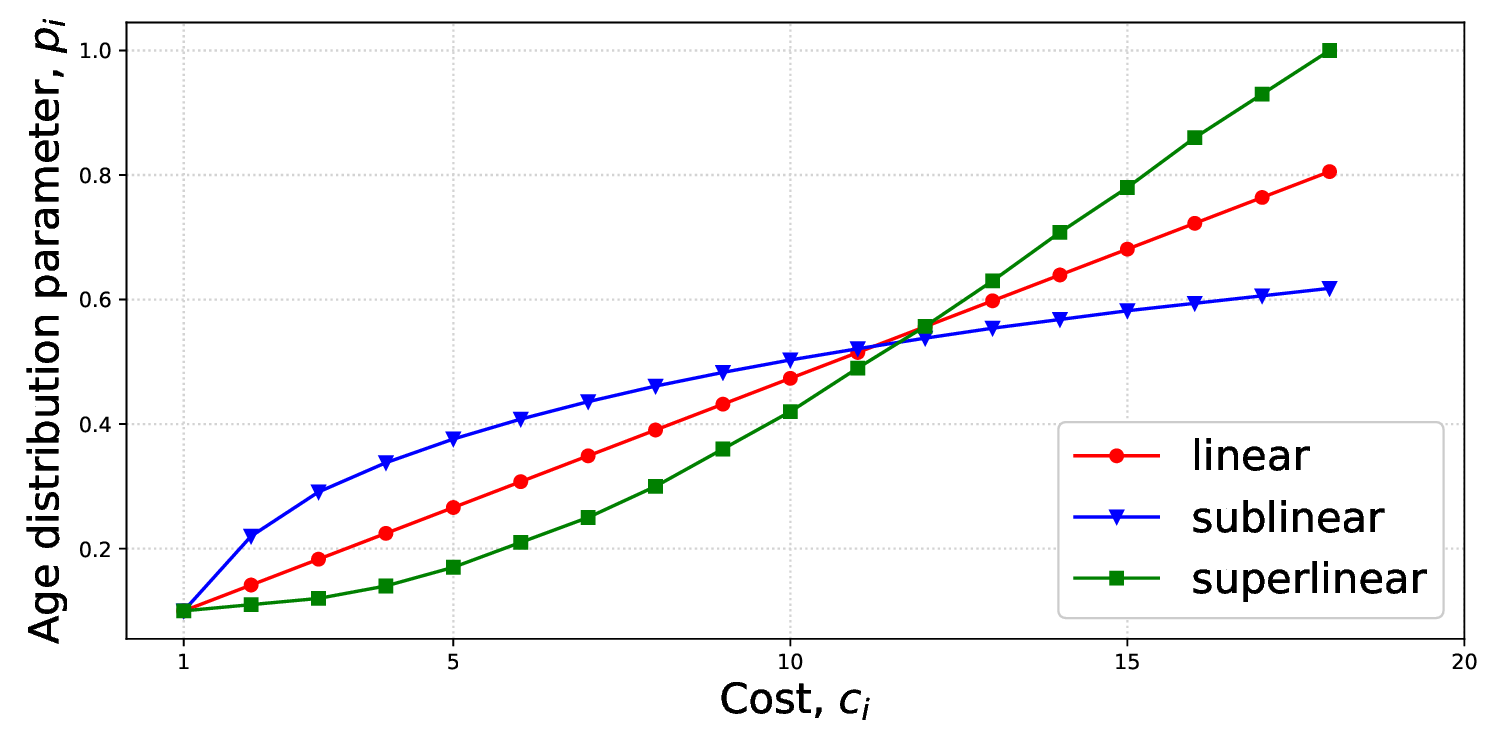}
	\vspace{-0.2cm}
	\caption{Cost-age distribution dependency for the sublinear cost scenario.}
	\label{fig:cost_reliability}%
	\vspace{-0.3cm}
\end{figure}

\subsection{Impact of different cost-reliability dependencies}

\begin{figure}[!t]	
	\centering
	\captionsetup{justification=centering}
	\includegraphics[width=.95\columnwidth]{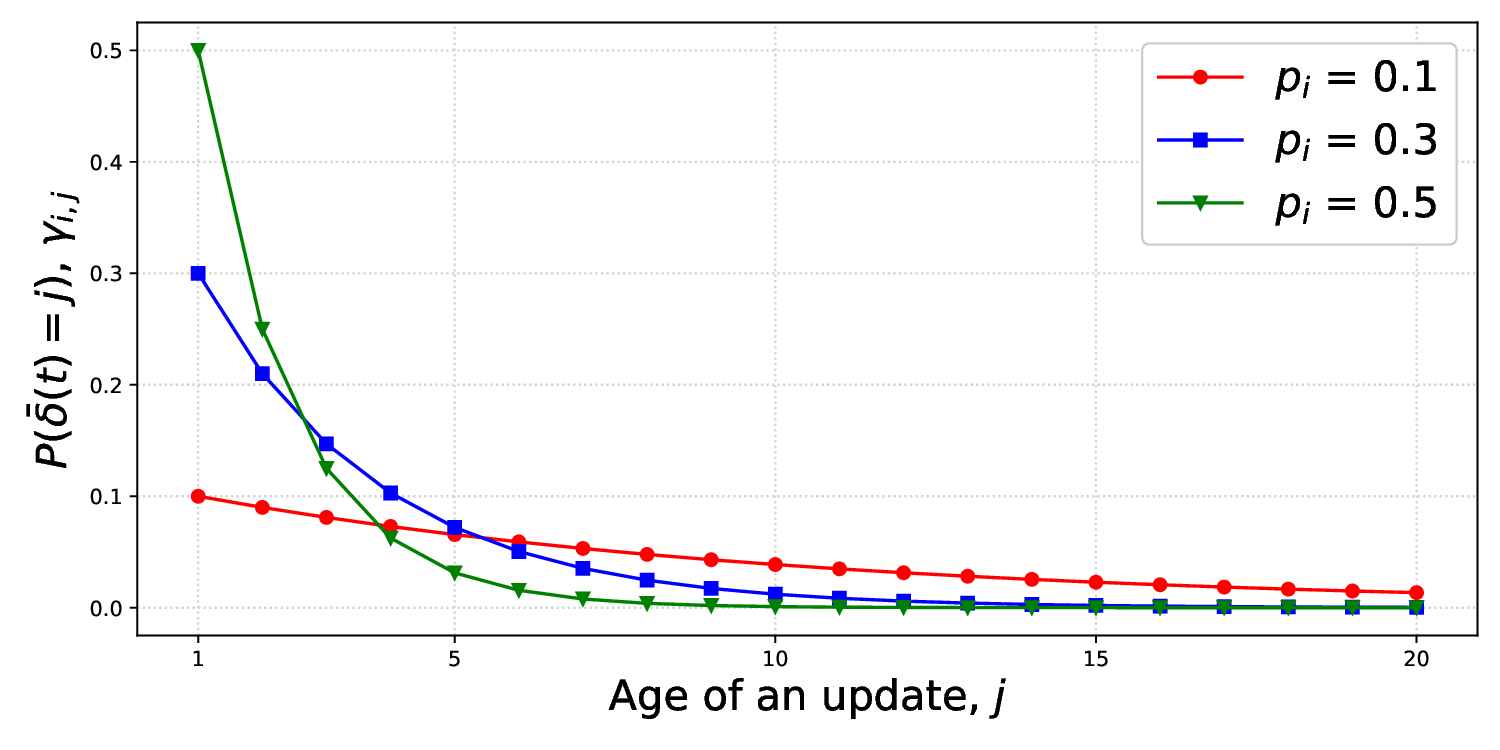}
	\vspace{-0.2cm}
	\caption{Geometric distribution of status updates for different $p_i$ parameters, where  $\bar{\delta}(t) \in [1, 20]$.}
	\label{fig:geometric_distribution}	
\end{figure}

Further, we evaluate the impact of different functions describing the cost-reliability dependencies. Similar with the cost function, in order to be able to perform a comparison we limit our attention to a specific class of age distributions from the sources. In particular, in our numerical analysis we assume that, for each $i$, $\gamma_{i, j}$ follows a geometric distribution with a different parameter $p_i$, as illustrated in Fig. \ref{fig:geometric_distribution}. This model also allows us to parametrize the distributions with a single parameter. Hence, the distribution of the age of the received status update, when the $i$-th information source is chosen, is given by:
\begin{equation}
	\gamma_{i, j} = Pr(\bar{\delta}(t)=j)=(1-p_i)^{j-1}p_i, \quad j = 1, 2, 3, ...., \beta-1
\end{equation}

Since we consider that packets with age higher than $\delta_{\rm max}$ have the same utility, we limit the geometric distribution to $\delta_{\rm max}$. Additionally, $\gamma_{i, \beta} = Pr(\bar{\delta}(t)=\beta) = 1 {-} \sum_{j=1}^{\beta {-} 1} (1{-}p_i)^{j-1}p_i$ for every $i$. As stated earlier, we expect to receive more fresh status updates from a more costly source, at least on average. To quantify such a relation, we consider the following general functional choices to relate $p_i \in [0, 1]$ with $c_i$ (Fig. \ref{fig:cost_reliability}):
\begin{equation}
	\text{Sublinear:} \quad p_i = k_{sub} \cdot c_i^2 ,	
	\label{eq:sub}
\end{equation}
\begin{equation}
	\text{Linear:} \quad p_i = k_{lin} \cdot c_i,
	\label{eq:lin}
\end{equation}
\begin{equation}
	\text{Superlinear:}  \quad  p_i =k_{sup} \cdot  \log_2 c_i,
	\label{eq:sup}
	\vspace{-0.2cm}
\end{equation}
where $k_{sub}$, $k_{lin}$, and $k_{sup}$ are chosen such that the average system parameters of age distribution ($p = \gamma_i$) is the same, i.e., $\frac{1}{n}\sum_{i=1}^n p_i$ is equal for the sublinear, linear and superlinear scenarios.

Once again, we would like to emphasize that our model and solution tools apply to arbitrary cost and age distributions, and these choices are made just to be able to observe the impact of three possible dependencies on the performance. 


\subsection{Results}

Default system parameters common to all the simulations are presented in Table \ref{tab:parameters}. The efficiency of the optimal and aggressive policies is verified via simulation runs over $T = 5000$ time slots, averaged over $M = 1000$  simulations. To demonstrate the results we plot the  AoI averaged over all times  $t = \overline{1,T}$. 

\begin{table}[t]
	\caption{Default System Parameters}
	\label{tab:parameters}
	\centering 
	\normalsize
	\small
	\begin{tabular}[h]{|p{7.0cm}|c|}
		\hline
		{\bf Parameters} & {\bf Values}\\
		\hline
		Battery capacity, $B$ & $20$ \\
		\hline
		AoI values of received updates, $[\alpha, \beta]$ & $[1,20]$ \\
		\hline
		Amount of harvested energy per time slot, $\{0, \bar{e}\}$ & $\{0, 3\}$ \\
		\hline
		Number of sources, $n$ & $8$ \\
		\hline
		Cost range, $[c_{\rm min}, c_{\rm max}]$ & $[1, 19]$ \\
		\hline
		Maximum AoI, $\delta_{\rm max}$ & $30$ \\
		\hline
		\end{tabular}
\end{table}

 \begin{figure*}[!t]%
 	\centering
	\stackunder[5pt]{\includegraphics[width=4in,height=.32in]{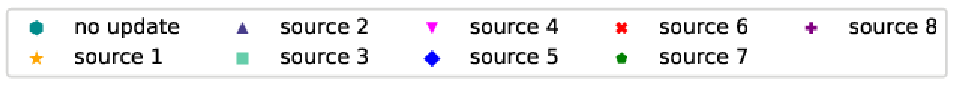}}{}
	\vspace{-0.3cm}
	
	\subfigure[$C$: superlinear; $\gamma$: superlinear]{%
		\includegraphics[trim=0mm 10mm 0mm 0mm, width=0.242\textwidth]{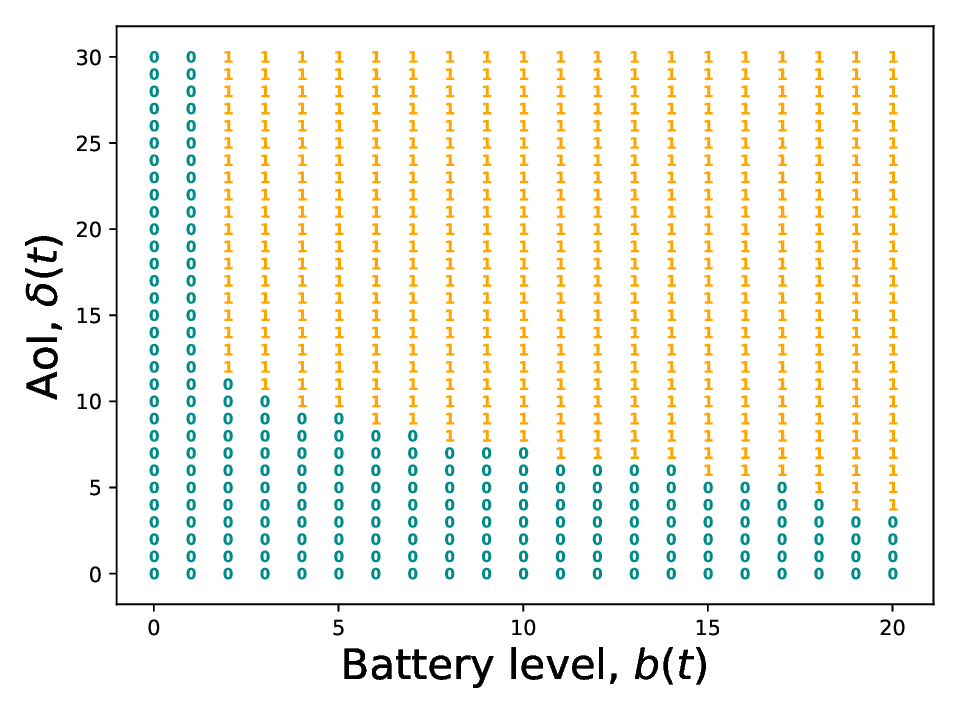}
		\label{fig:02_sup_sup}%
	}
	\subfigure[$C$: superlinear; $\gamma$: linear]{%
		\includegraphics[width=0.242\textwidth, trim=0mm 10mm 0mm 0mm]{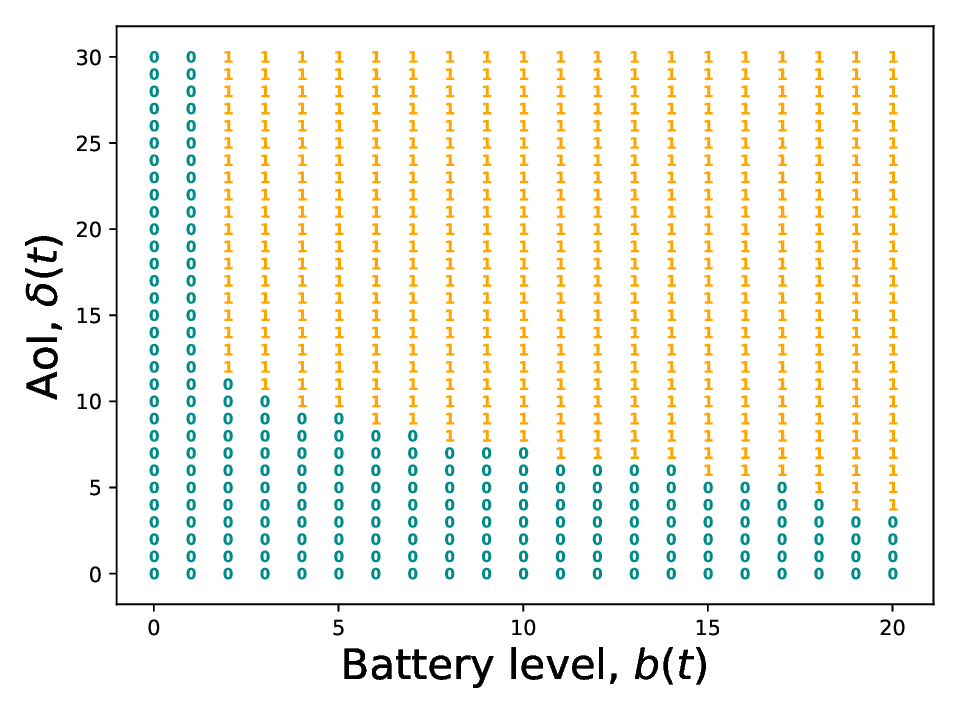}
		\label{fig:02_sup_lin}%
	}
	\subfigure[$C$: superlinear; $\gamma$: sublinear]{%
		\includegraphics[width=0.242\textwidth, trim=0mm 10mm 0mm 0mm]{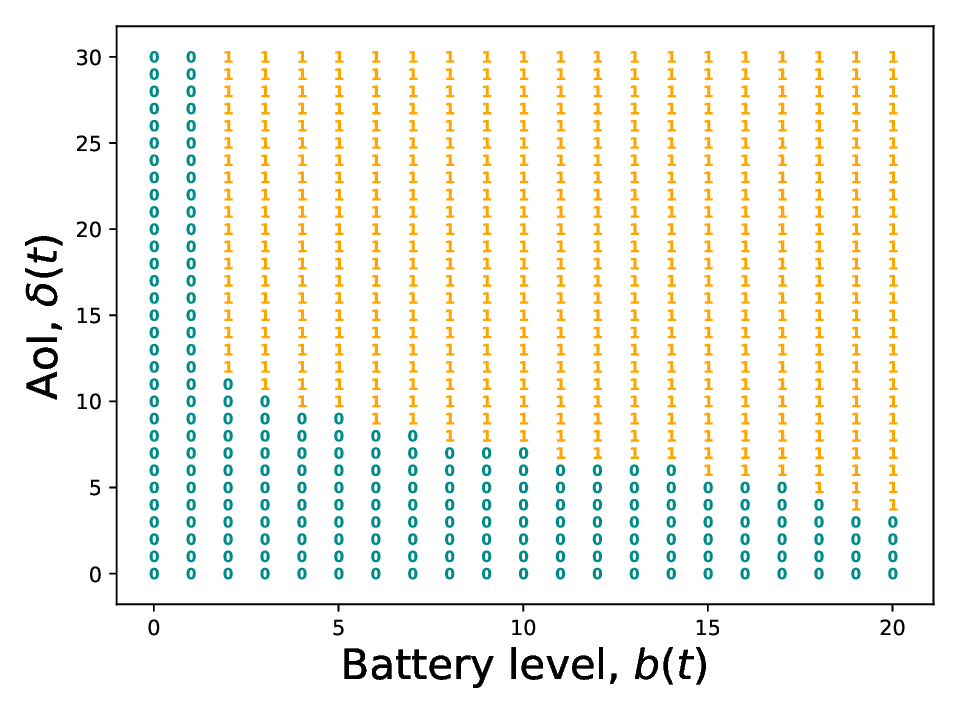}
		\label{fig:02_sup_sub}%
	}
	\vspace{-5pt}
	\subfigure[$C$: linear; $\gamma$: superlinear]{%
		\includegraphics[width=0.242\textwidth, trim=0mm 10mm 0mm 0mm]{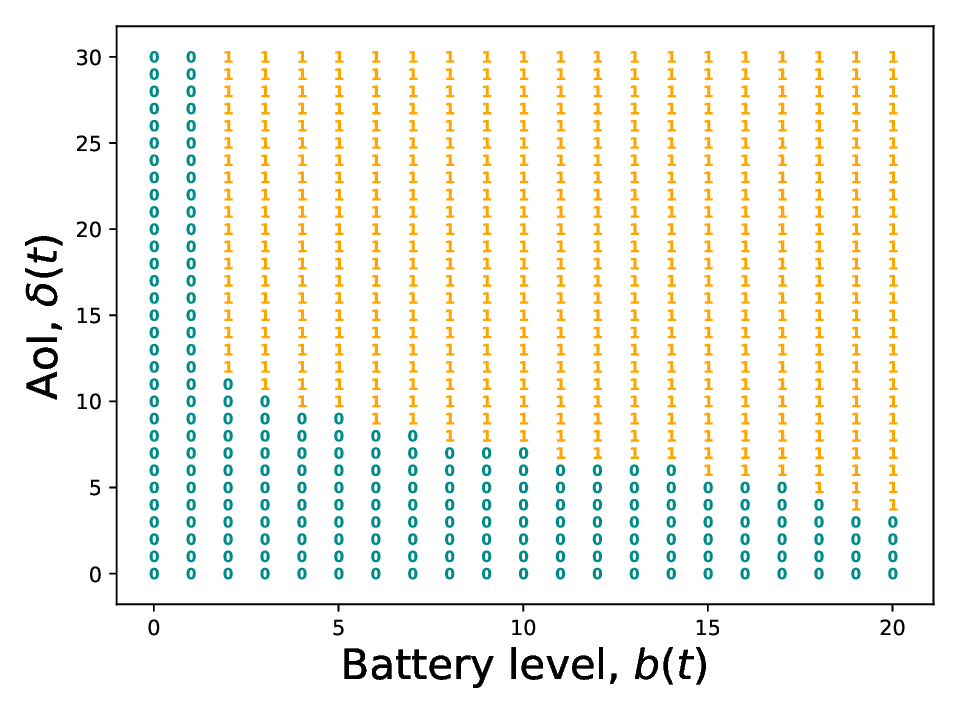}
		\label{fig:02_lin_sup}%
	}
	\subfigure[$C$: linear; $\gamma$: linear]{%
		\includegraphics[width=0.242\textwidth, trim=0mm 10mm 0mm 0mm]{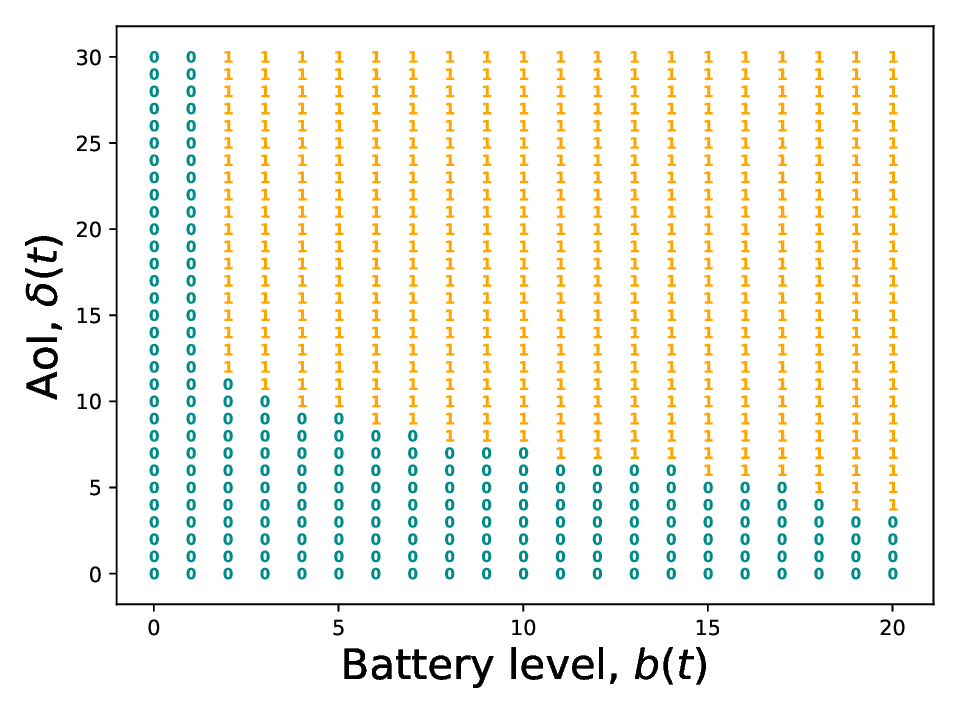}
		\label{fig:02_lin_lin}%
	}
	\subfigure[$C$: linear; $\gamma$: sublinear]{%
		\includegraphics[width=0.242\textwidth, trim=0mm 10mm 0mm 0mm]{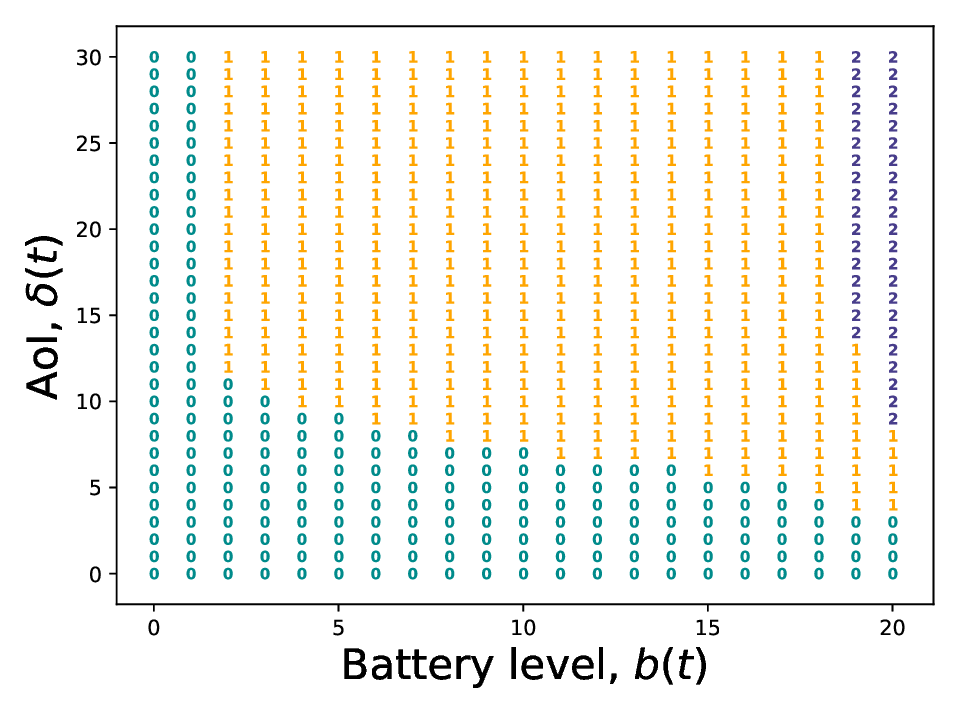}
		\label{fig:02_lin_sub}%
	}
	\vspace{-5pt}
	\subfigure[$C$: sublinear; $\gamma$: superlinear]{%
		\includegraphics[width=0.242\textwidth, trim=0mm 10mm 0mm 0mm]{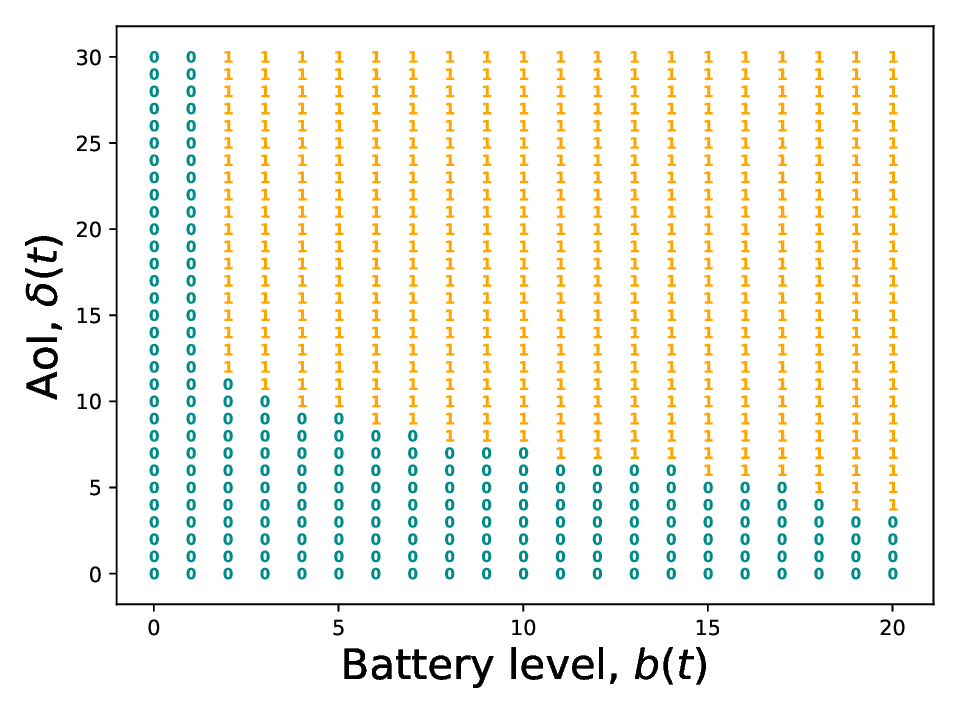}
		\label{fig:02_sub_sup}%
	}
	\subfigure[$C$: sublinear; $\gamma$: linear]{%
	\includegraphics[width=0.242\textwidth, trim=0mm 10mm 0mm 0mm]{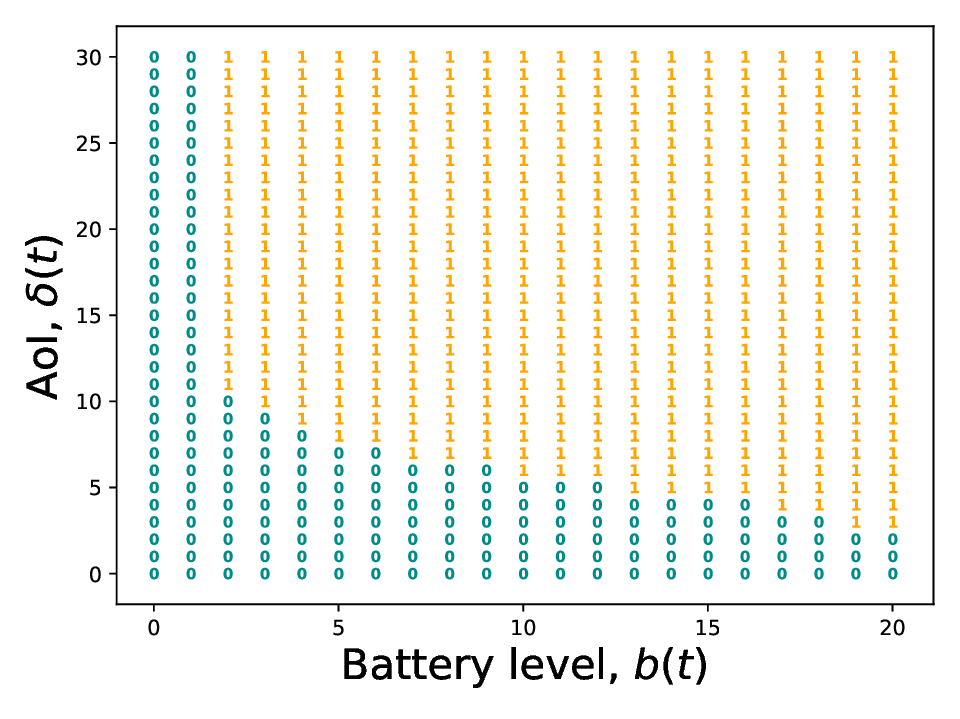}
	\label{fig:02_sub_lin}%
	}
	\subfigure[$C$: sublinear; $\gamma$: sublinear]{%
		\includegraphics[width=0.242\textwidth, trim=0mm 10mm 0mm 0mm]{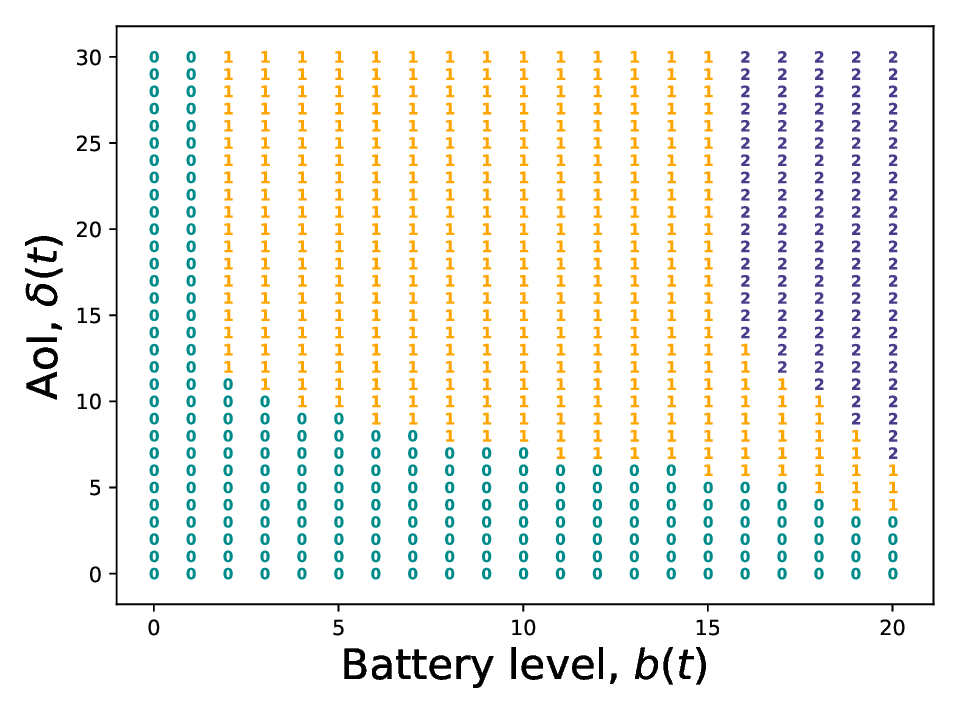}
		\label{fig:02_sub_sub}%
	}
	\caption{Illustration of the optimal policy for different energy cost/ age distribution combinations for EH rate $\lambda = 0.2$.}
	\label{fig:02}
\end{figure*}

\begin{figure*}[!t]%
 	\centering
	\stackunder[5pt]{\includegraphics[width=4in,height=.32in]{./Figures/legend.eps}}{}
	\vspace{-0.3cm}

 	\subfigure[$C$: superlinear; $\gamma$: superlinear]{%
 		\includegraphics[width=0.242\textwidth, trim=0mm 10mm 0mm 0mm]{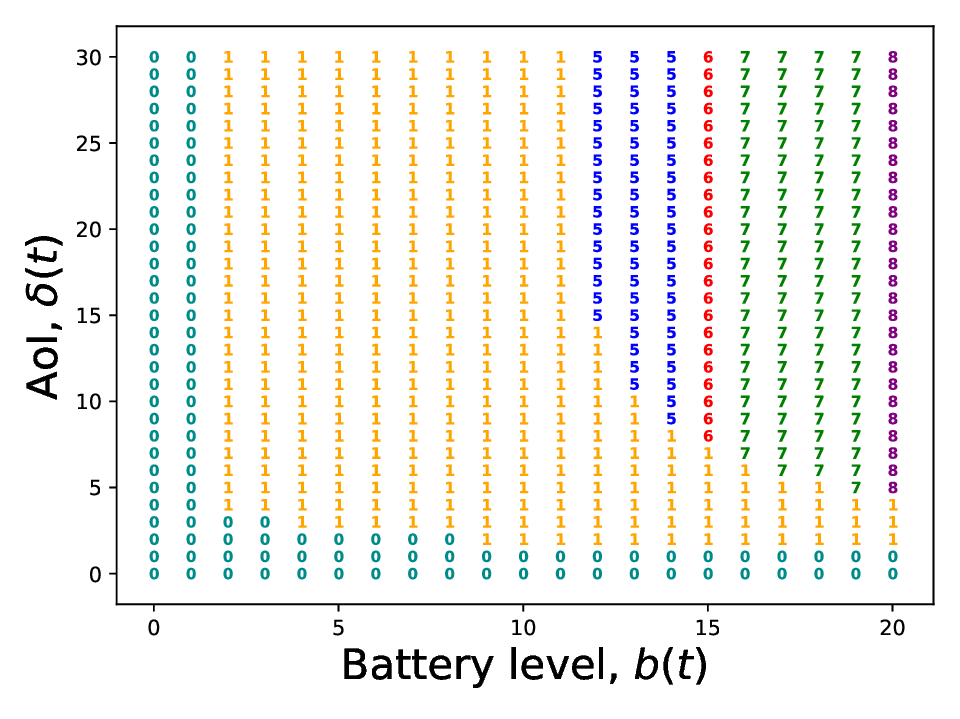}
 		\label{fig:08_sup_sup}%
 	}
 	\subfigure[$C$: superlinear; $\gamma$: linear]{%
 		\includegraphics[width=0.242\textwidth, trim=0mm 10mm 0mm 0mm]{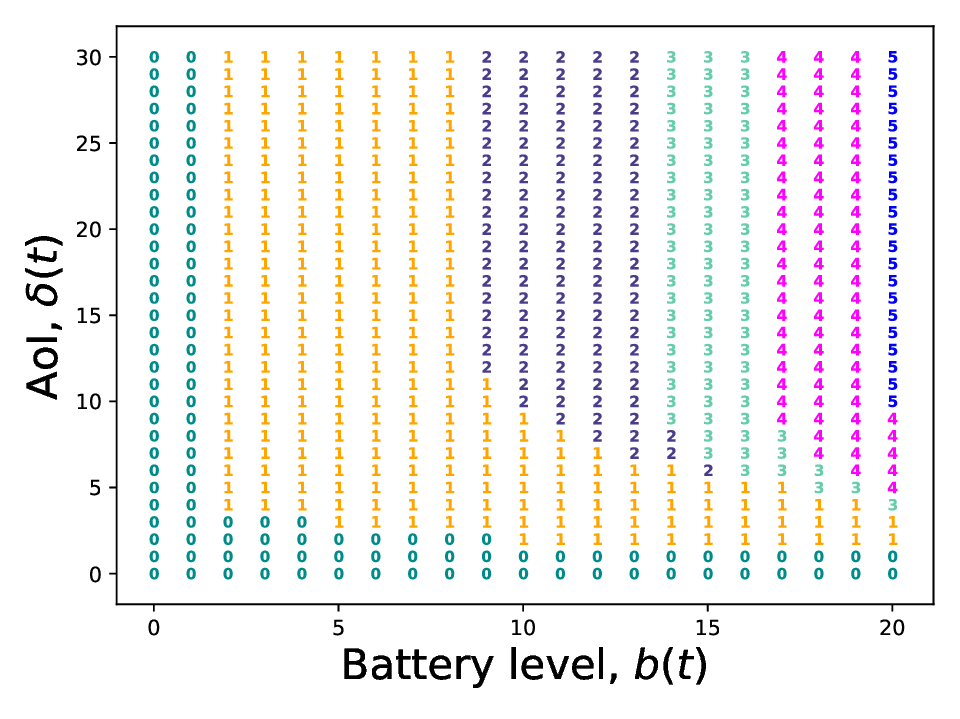}
 		\label{fig:08_sup_lin}%
 	}
 	\subfigure[$C$: superlinear; $\gamma$: sublinear]{%
 		\includegraphics[width=0.242\textwidth, trim=0mm 10mm 0mm 0mm]{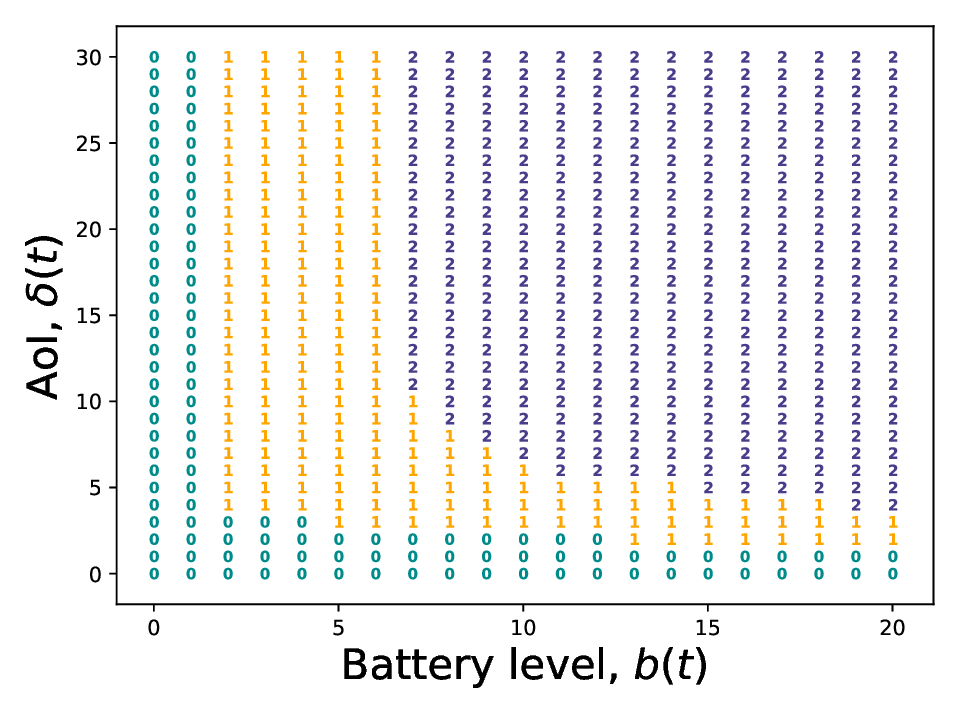}
 		\label{fig:08_sup_sub}%
 	}
 	\vspace{-5pt}
 	\subfigure[$C$: linear; $\gamma$: superlinear]{%
 		\includegraphics[width=0.242\textwidth, trim=0mm 10mm 0mm 0mm]{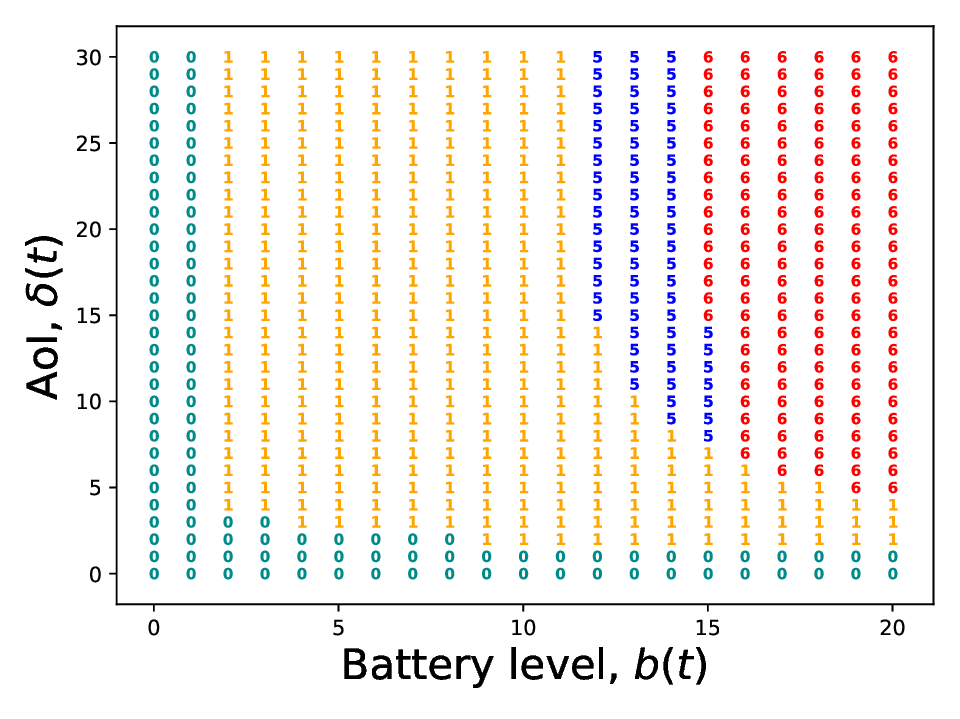}
 		\label{fig:08_lin_sup}%
 	}
 	\subfigure[$C$: linear; $\gamma$: linear]{%
 		\includegraphics[width=0.242\textwidth, trim=0mm 10mm 0mm 0mm]{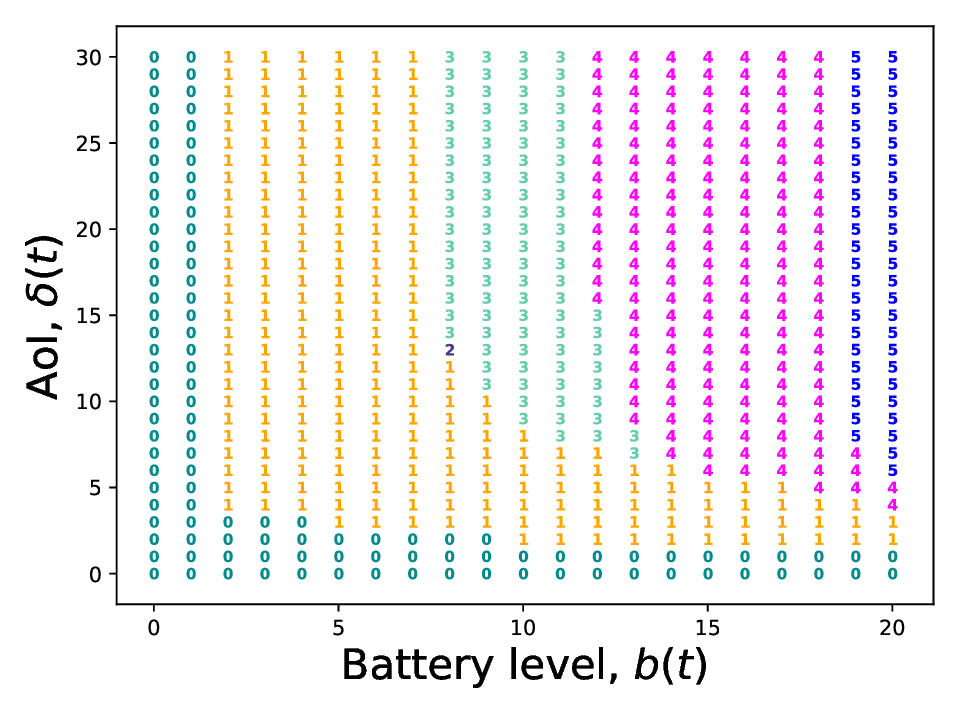}
 		\label{fig:08_lin_lin}%
 	}
 	\subfigure[$C$: linear; $\gamma$: sublinear]{%
 		\includegraphics[width=0.242\textwidth, trim=0mm 10mm 0mm 0mm]{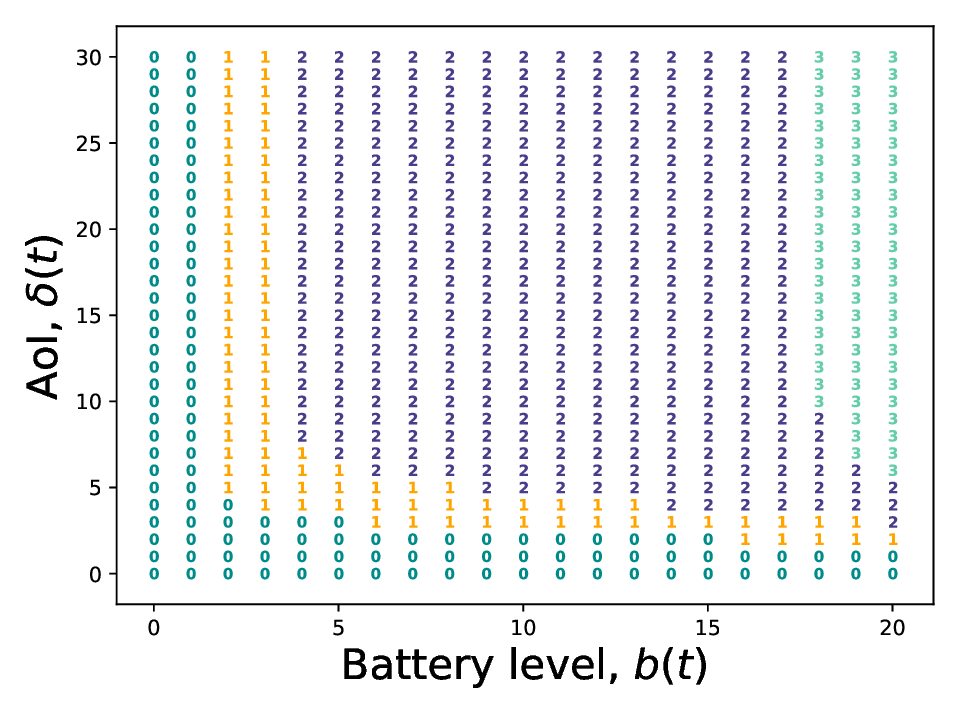}
 		\label{fig:08_lin_sub}%
 	}
 	\vspace{-5pt}
 	\subfigure[$C$: sublinear; $\gamma$: superlinear]{%
 		\includegraphics[width=0.242\textwidth, trim=0mm 10mm 0mm 0mm]{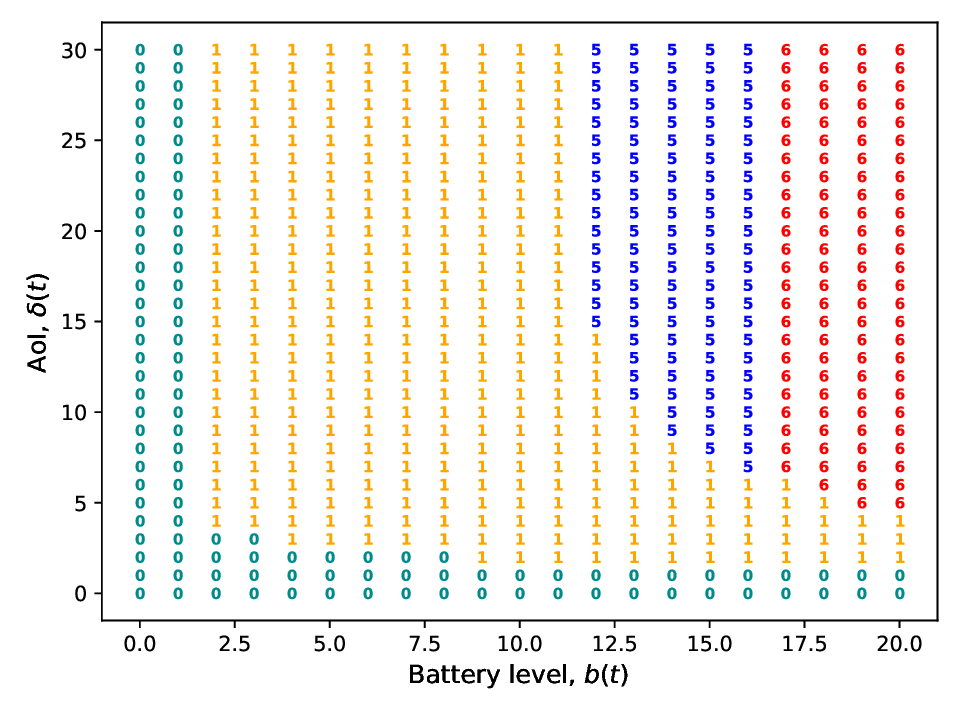}
 		\label{fig:08_sub_sup}%
 	}
 	\subfigure[$C$: sublinear; $\gamma$: linear]{%
 		\includegraphics[width=0.242\textwidth, trim=0mm 10mm 0mm 0mm]{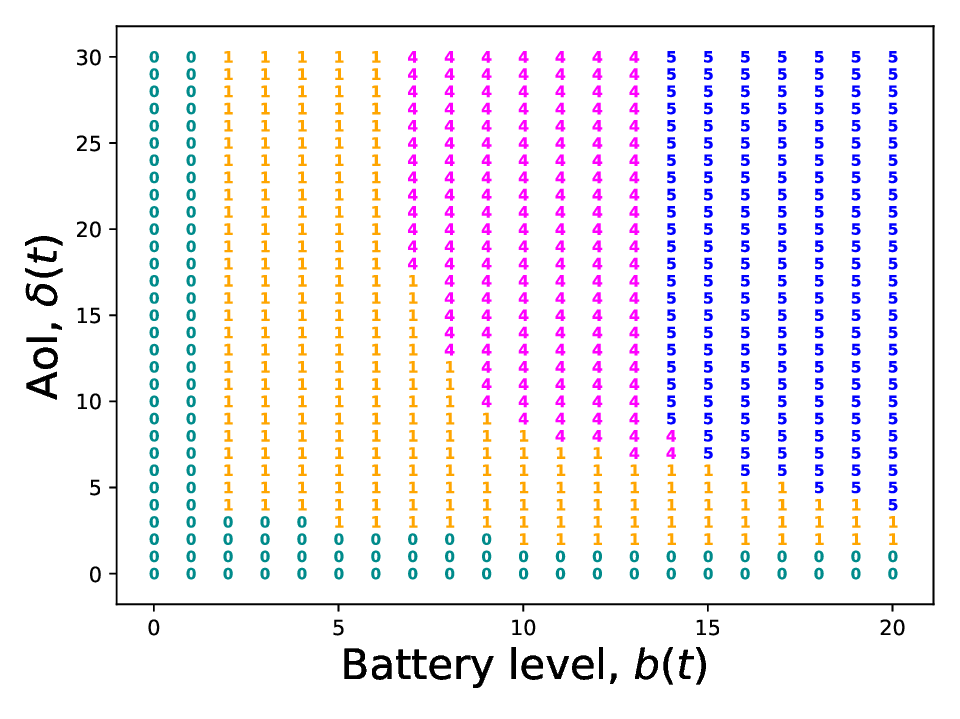}
 		\label{fig:08_sub_lin}%
 	}
 	\subfigure[$C$: sublinear; $\gamma$: sublinear]{%
 		\includegraphics[width=0.242\textwidth, trim=0mm 10mm 0mm 0mm]{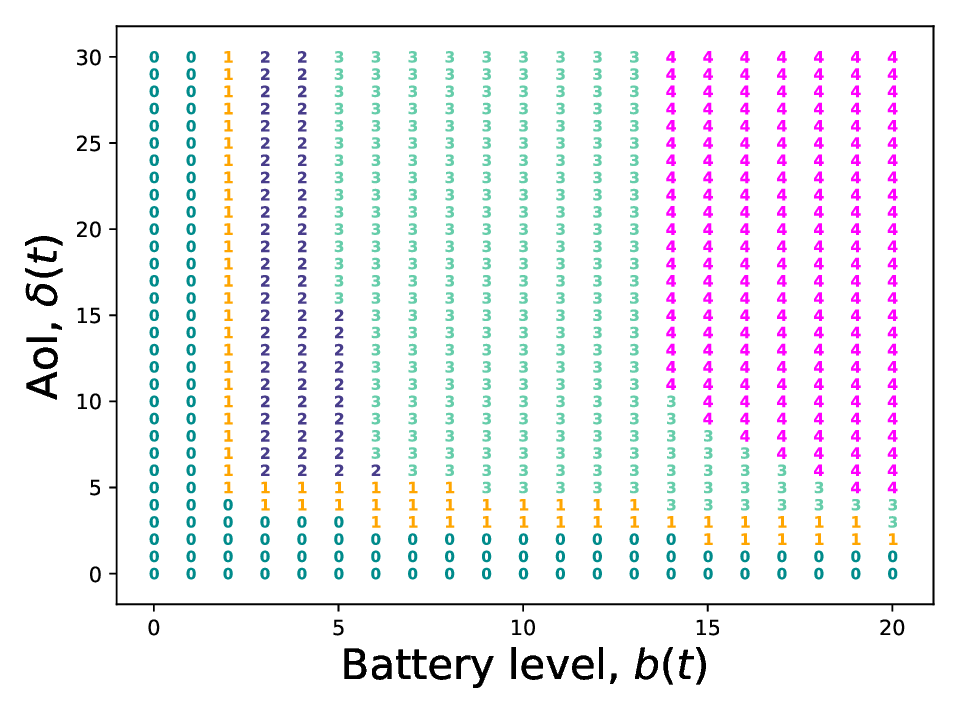}
 		\label{fig:08_sub_sub}%
 	}
 	\caption{Illustration of the optimal policy for different energy cost/ age distribution combinations for EH rate $\lambda = 0.6$.}
 	\label{fig:08}
 \end{figure*}

The optimal solutions for different values of EH rates and cost-age distribution combinations are presented in Figs.\ \ref{fig:02} and \ref{fig:08}. Both figures show the 9 possible combinations of cost and age distribution each taking values denoted by {superlinear, linear, sublinear} as in \eq{eq:sub} - \eq{eq:sup}, see also Figs.\ 	\ref{fig:rank_cost} and \ref{fig:cost_reliability}.

We set $\lambda = 0.2$ in Fig.\ \ref{fig:02} and $\lambda = 0.6$ in Fig.\ \ref{fig:08}. Each of the 9 subfigures shows the optimal policy depending on the system state. 
In both cases, $n=8$ sources are considered, so the optimal policy chooses among 9 possible actions including ``no update'' (i.e., to stay idle).

Fig.\ \ref{fig:02} shows that when the EH rate is low, i.e., $\lambda = 0.2$, the monitoring node requests a status update only from the cheapest sources, i.e., sources 1, 2. Notably, the result is similar for all the combinations of cost and age distributions. In particular, the \emph{activity region}, i.e.,  the set of states in which the monitoring node is actively requesting updates, remains the same. The activity region requires  that both  battery level and AoI are high enough to request an update. 

For low values of AoI, the monitoring node never requests a status update, since the information is still fresh. Also, for  low values of the battery level a status update cannot be afforded. However, differently from the aggressive approach, where an update is always requested if there is enough energy in the battery, the optimal policy, in contrast, conserves energy if the AoI is sufficiently low. This leads to an \emph{energy saving region} for $\delta(t) \in [0, \delta_u(b(t))]$, where $\delta_u(b(t))$ is the highest AoI value for which no update is requested. The value of $\delta_u(b(t))$ decreases with $b(t)$, because at high battery levels the monitoring node can be more relaxed in status update requests. This trend applies for all cost-age distribution combinations in the same way. The only difference appears when the dependency of the parameters of the age distribution is sublinear, due to the fact that the more expensive source 2 is significantly more reliable, and therefore, worth using at higher energy levels.  However, this also depends on the cost of source 2; if the cost dependence is also sublinear then source 2 is employed instead of source 1 for lower values of $b(t)$. 
 
 The common aspect of all the 9 subplots  in Fig. \ref{fig:02} is that only cheap sources are used when the energy arrivals are scarce. In contrast, Fig. \ref{fig:08} shows more variations in the usage of different sources for $\lambda = 0.6$. Here, similarly to Fig. \ref{fig:02}, the same 9 cases are considered in the respective subfigures. For $\lambda = 0.6$, multiple sources are used depending not only on the values of $b(t)$ and $\delta(t)$, but also on the cost and age distribution combinations. Even though the \emph{activity region} is approximately the same, it is split differently among multiple sources, and not necessarily only the cheapest ones. 
 In particular, the baseline case where cost and dependency of age distribution parameters are both linear (Fig. \ref{fig:08_lin_lin}) demonstrates that a wide array of sources from 1 to 5 (i.e., the 5 cheapest ones) are usedThe higher $b(t)$ and/or $\delta(t)$, the higher the index of the source used for the update.
 
  If we change the cost from superlinear to sublinear (Fig. \ref{fig:08_sup_lin}, \ref{fig:08_lin_lin}, \ref{fig:08_sub_lin}), we see that, within the \emph{activity region}, source 1 is used more or less consistently in all the 3 cases, but the patterns if other sources change, with intermediate sources becoming more widely used if the cost is sublinear. This trend is generally true if we read the subfigures from top to bottom. Based on the structural difference of the optimal solutions, the choice of the specific function is less important compared to its characteristics in terms of concavity/convexity. 
  
  Conversely, if we change the dependency of age distribution parameters (Fig, \ref{fig:08_lin_sup}, \ref{fig:08_lin_lin}, \ref{fig:08_lin_sub}), the cheaper sources are used more often, and their usage happens at lower battery levels, i.e., their region shifts towards left. This trend is also generally true if we read the subfigures from left to right. 

\vspace{-0.1cm}
\subsection{Discussion}

In this section, we prove some structural property of the optimal policy, in particular, an existence of a threshold effect on the battery level $b$ (but notably, not on $\delta$). 

\begin{theorem}
	If the AoI is unlimited (or, its maximum value is sufficiently high) 
	then the optimal policy has an AoI-threshold-based behavior that holds for any value of  $b$.
\end{theorem}
This means that if we focus on a given $b$, the optimal policy depends on the AoI $\delta$ such that:
\begin{itemize}
	\item a given subset of $k(b)$ sources is used, denoted by $\sigma_1(b), \sigma_2(b), \dots \sigma_{k(b)}(b) \in [N]$,
	\item exactly $k(b)$ threshold values for the AoI $\delta$, denoted by $\vartheta_{1}(b), \vartheta_{2}(b), \dots,\vartheta_{k(b)}(b)$ can be defined, in strictly increasing order (i.e., $\vartheta_{j}(b) < \vartheta_{j+1}(b)$ for every $j \in [N{-}1]$), so that source $\sigma_j(b)$ is used only when $\vartheta_j(b) \leq \delta < \vartheta_{j+1}(b)$ for $j {\in} [N{-}1]$, and the last source $\sigma_{k(b)}$ is used for $\delta \geq \vartheta_{k(b)}(b)$, while no update is attempted if $\delta < \vartheta_1(b)$.
\end{itemize}

This threshold-based character of the optimal policy under the aforementioned conditions, can be proven through the following two lemmas.
\begin{lemma}
	\label{lemma:2}
	A system with $n \geq 2$ sources has an optimal AoI-threshold-based activation $\vartheta_{1}(b)$ for all values of $e$, meaning the optimal policy is to stay idle (action $a_0$) when $\delta < \vartheta_{1}(b)$, and conversely action $a_0$ is suboptimal if $\delta \geq \vartheta_{1}(b)$.
\end{lemma}%
\begin{proof}
	The details are reported in Appendix \ref{sec:proof_1}. 
\end{proof}

Following this Lemma, one can see that  if $\delta \geq \vartheta_1(b)$ it is convenient to update but it is not knowns from which source. We need to obtain a full AoI-threshold-based structure as required by the theorem to show that, if a given $b$ is considered and $\delta$ is increased from $\vartheta_1(b)$, there are other turning points $\vartheta_2(b), \vartheta_3(b), \dots$ such that for $\delta \geq \vartheta_j(b)$ the optimal action switches from $s_{j}(b)$ to $s_{j+1}(b)$ and \emph{never reverts back} to $s_{j}(b)$ after that point.
This is shown through this last Lemma.

\begin{lemma}
	\label{theorem_2}
		Consider a given value of $b$ and two different sources $i$
		and $j$, whose associated
		update actions are $a_i$ and $a_j$, respectively. If $a_i$ is
		preferable over $a_j$ when the battery level is $b$  and the AoI value is $\delta_1$, where $\delta_1 \geq  v_1(b) $, and the reverse happens (i.e.,
		$a_j$ is preferable over $a_i$) for battery level $b$ and AoI equal to
		$\delta_2 > \delta_1$, then $a_j$ is necessarily better than $a_i$ for
		battery level $b$ and all AoI values $\delta > \delta_2$.
\end{lemma}

The proof is provided in Appendix  \ref{sec:proof_2}.
We remark that the theorem proves an AoI-threshold-based behavior, but in general we do not have
a similar behavior in the battery level $b$, as discussed in the following counterexample.




{\bf Counterexample}. Consider a scenario with two information sources, where $\delta_{\max} = 1$, and $\alpha =0$ $\beta=1$. The  two information sources $1$ and $2$ have the following characteristics:  $0 < c_1 <c_2 < B$, $p_2 \gg p_1$. 

Compare the time-average AoI for $T = 2$ for the following sequences of actions performed by the monitoring node:

\textit{Sequence 1:} (use $1$; use $1$), resulting in $\bar{\delta} {=} 1{-}p_1 {+} \frac{(1{-}p_1)^2}{2}$

\textit{Sequence 2:} (do not transmit; use $2$), resulting in $\bar{\delta} {=}  1 {+} \frac{1{-}p_2}{2}$

If $p_2 > 4p_1 - p_1^2$, then sequence $2$ is preferrable over sequence $1$. However, sequence $2$ may not be available if using source $2$ is too expensive for the current battery state. In other words, depending on the current energy state and energy arrivals, it may be more convenient to just use the cheap source or to wait in order to enable the expensive source in the subsequent time slot. Formally, this happens if $b(0) - c_1 + \lambda h > c_2$. Hence, we proved that with an increase in the state of the battery at time $0$, the minimization of AoI can imply to use a \emph{cheaper} source at time $1$ (in this specific example, no source at all), which contradicts the monotonicity of the source index in the battery level.

In Fig. \ref{fig:coubterexample1}, we demonstrate the effect of limiting $N$ on the structure of the optimal solution. We considered a system with two sources with $p_1 = 0.2$, $p_2 = 0.99$, $c_1 = 5$, $c_2 = 11$. The AoI distribution is kept geometric in range  $[\alpha, \beta] = [0, 20]$. The energy-harvesting process is given with parameters $\lambda = 0.9$, $\bar{e} = 5$.  We observe the energy saving region that occurs under this particular combination of parameters.

Another visual counterexample is also graphically presented in Fig. 7, where we adopt the default settings from Table I, and consider two sources such that $p_2 \gg p_1$, $\lambda = 0.9$.  We also increased the value of parameter $\bar{e}$ so that the energy buffer can recover fast. The costs of sources are set as follows $c_1 = 1, c_2 = 16$. The distribution of AoI is preserved as geometric in range  $[\alpha, \beta] = [1,20]$. In this setup, the optimal policy is not threshold-based with respect to the battery level.

\subsection{Performance comparison}

  \begin{figure}[!t]	
	
	\centering
	\includegraphics[width=0.72\columnwidth]{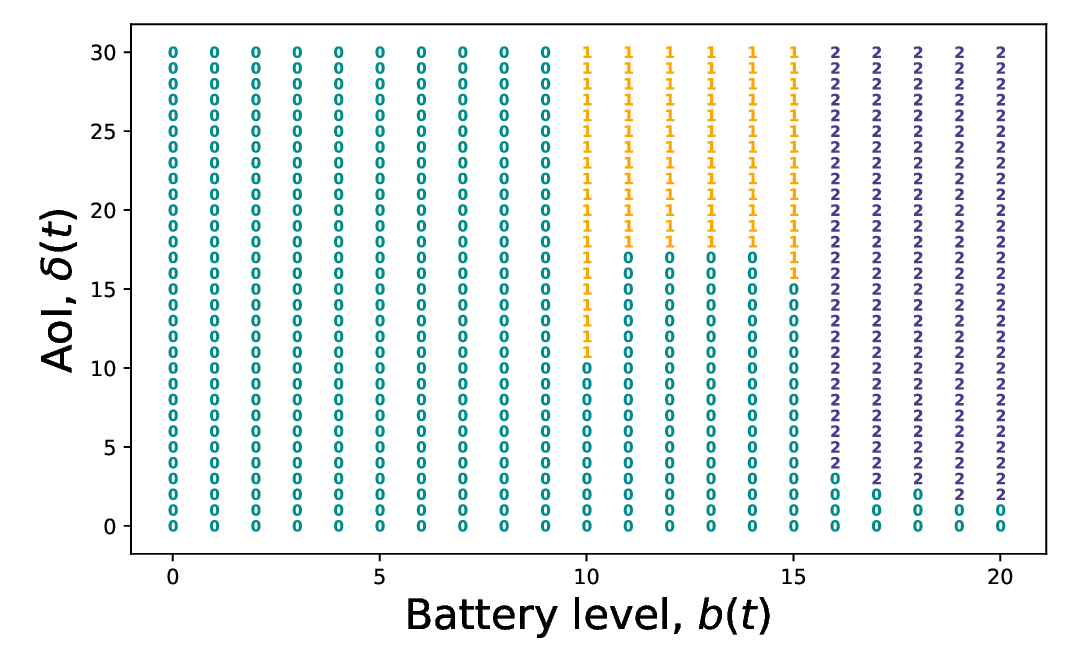}
	\caption{Optimal solution ($p_2= 1$, $p_1 = 0.1$, $\lambda = 0.9$).}
	\label{fig:pocket}	
\end{figure}
\begin{figure}[!t]	
	\centering
	\includegraphics[width=0.72\columnwidth]{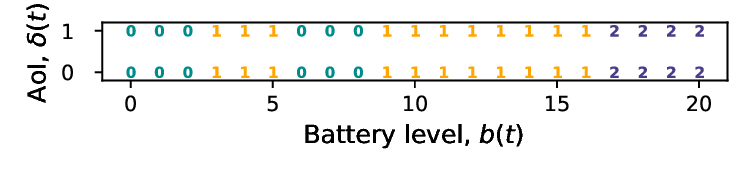}
	\caption{Optimal solution: $N=1$, $p_2 = 0.99$, $c_1 = 5$, $c_2 = 11$, $\lambda = 0.9$, $\bar{e} = 5$}
	\label{fig:coubterexample1}	
\end{figure}

\begin{figure}[!t]	
	\centering
	\includegraphics[width=0.79\columnwidth]{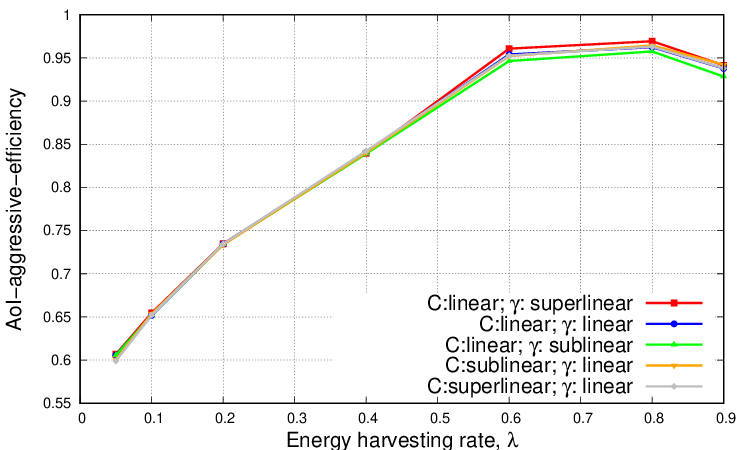}
	\caption{Rate between the average AoI obtained by the optimal and aggressive policies as a function of the EH rate.}
	\label{fig:average_aoi}	
\end{figure}

To understand the potential benefits of the optimal policy, we compare it with the aggressive policy as a benchmark. 
In Fig. \ref{fig:average_aoi} we plot the relative gain over the aggressive policy vs the EH rate $\lambda$, where the AoI-aggressive-efficiency in the y-axis is defined as the ratio between the average AoI obtained by the optimal strategy to the one obtained by the aggressive policy.
 For the sake of brevity, five cost and age distribution combinations are considered. High (close to 1) AoI-aggressive-efficiency implies that the aggressive policy is quite efficient, and benefit of using the more computationally demanding VI framework is limited. Despite some differences in the structure of the optimal solution, the resulting AoI-aggressive-efficiency has similar values for all cost-age distribution combinations. The AoI-efficiency-rate increases with $\lambda$,  meaning that the difference in performance between optimal and aggressive policy vanishes at high $\lambda$. In particular, for $\lambda > 0.5$ the AoI-aggressive-efficiency saturates above 0.90. We can conclude, that if the energy arrivals are relatively stable, the benefits of optimization is rather limited. 
 On the other hand, for low values of $\lambda$ the optimization of the update policy is much more relevant, which follows the intuition. 
Yet, when $\lambda$ is very low, the role of multiple sources is minimal, and only the cheapest sources are used (see  Fig. \ref{fig:02}).

\begin{figure}[!t]	
	\centering
	\includegraphics[width=0.91\columnwidth]{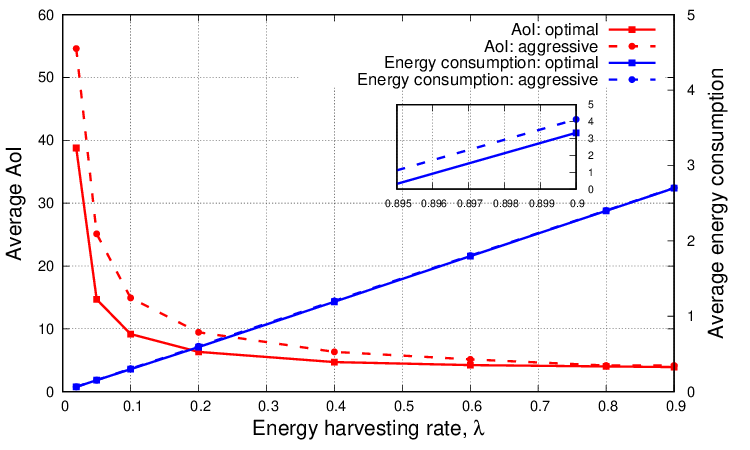}
	\caption{EH rate, $\lambda$ vs average AoI and average energy consumption.}
	\label{fig:aoi_energy}	
\end{figure}

In Fig. \ref{fig:aoi_energy}, we plot the average AoI and the average energy consumption vs.\  the EH rate. As one would expect, the average energy consumption increases with $\lambda$, while the average AoI decreases. We observe that the two policies have almost identical energy consumption, for low $\lambda$ values, although the optimal policy provides significantly lower average AoI performance. Also, the energy consumption of the aggressive policy saturates at high $\lambda$ values, while that of the optimal policy continue to increase linearly.

\begin{figure}[!t]	
	\centering
	\includegraphics[width=0.91\columnwidth]{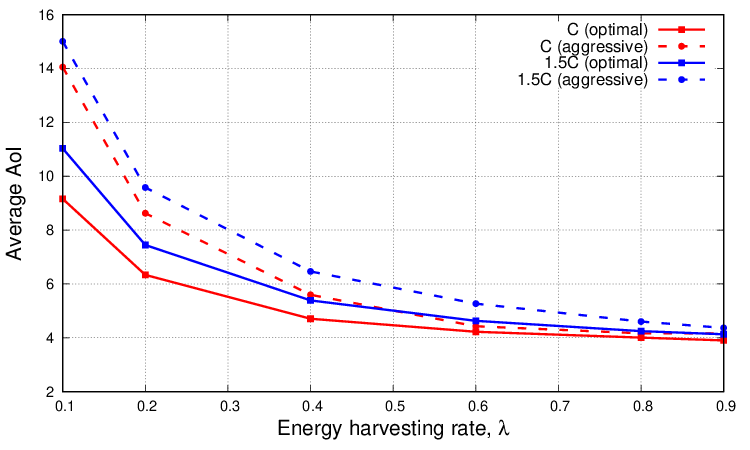}
	\caption{EH rate, $\lambda$, vs average AoI for linear cost-age distribution.}
	\label{fig:geometric_prop}	
\end{figure}

We also analyzed the average AoI when the average update cost is 50\% higher than the default case. To do so, we increased the cost of each source $1.5$ times ($1.5C$) and found the average AoI for the linear cost-age distribution case, as demonstrated in Fig. \ref{fig:geometric_prop}. With the increase of cost, the difference between the average AoI achieved by the optimal and aggressive policies reaches up to 15\%, if $\lambda$ is low. When $\lambda$ is high, the difference in performance is insignificant.

\subsection{Network size}

Further, we analyze how the number $n$ of information sources affects the performance for different values of EH rate, $\lambda$, and energy arrival units, $\bar{e}$. The analysis is performed for linear scaling of the costs of the devices, and a linear dependency between the cost and the number of the sources. 

We decrease the space of actions (or network size) in the following manner: first, we form the vector of size $n$ with costs $[c_1, c_2, ...., c_n]$, and derive the average AoI for $n$ information sources. Then, we reduce the network size by half at each step till we have only two information sources with  cost vector $[c_1, c_n]$, thus we obtain the average AoI for $n = 2, 4, 8, 16$. For $n = 12$, we randomly removed four sources. 

\begin{figure}[!t]	
	\centering
	\includegraphics[width=0.92\columnwidth]{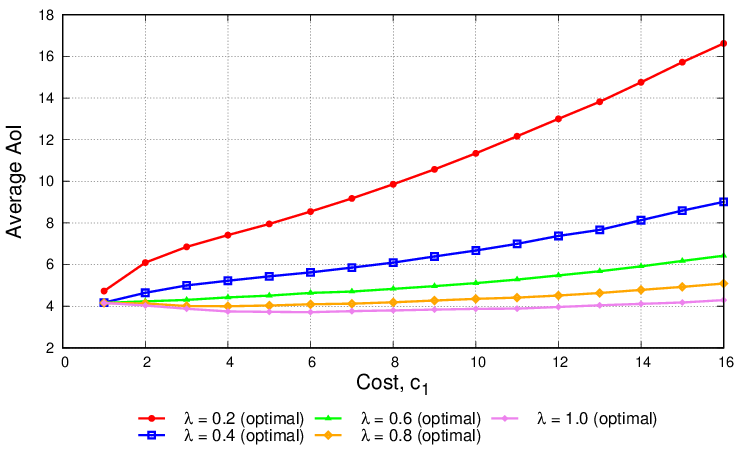}
	\caption{Optimal average AoI  vs. cost of an information source for different values EH rates, $\lambda$.}
	\label{fig:one_source}
\end{figure}

Firstly, we consider a system without sources diversity, i.e., with a single information source; and demonstrate the dependency of optimal average AoI and cost of that source. With the increase in the cost, the optimal average AoI increases, despite the fact that  with the increase of the cost, the probability to receive a fresh status update increases. Moreover, in the greatest extent, an increase in cost affects the performance in case of low frequency energy arrivals (see Fig.\ \ref{fig:one_source}). If the cost value is low, i.e., $c_1 = 1$, the performance for different values of $\lambda$ has minor variation. In particular, the performance is identical if $\lambda$ is sufficiently high ($\lambda \geq 0.4$). Although, when $\lambda = 0.2$ the optimal solution has a larger energy saving area, which is why we observe a larger gap in performance.

\begin{figure}[!t]	
	\centering
	\includegraphics[width=0.92\columnwidth]{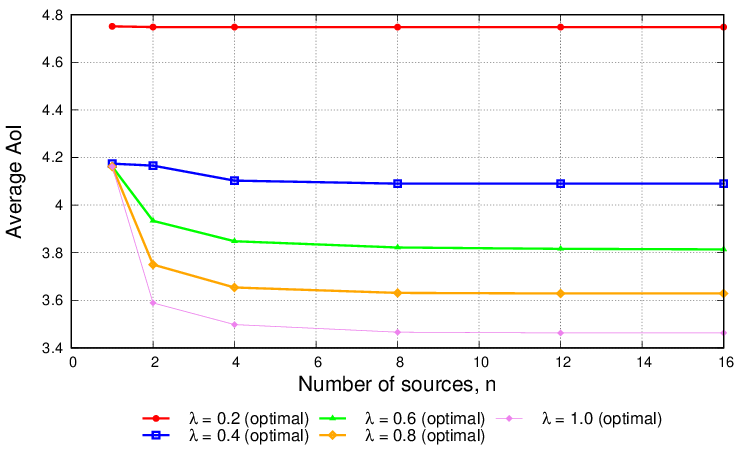}
	\caption{Optimal average AoI  vs. number of information sources for different values of EH rates, $\lambda$.}
	\label{fig:number_optimalAoI_lambda}
\end{figure}

With the increase in the number of information sources, the optimal average AoI has a tendency to decrease, but the curves eventually saturate when the number of sources reaches $n = 8$ (Fig.\ \ref{fig:number_optimalAoI_lambda}). The largest gain in performance is obtained when the system is of size $n = 2$. If the EH rate is low ($\lambda = 0.2$ in Fig.\ \ref{fig:number_optimalAoI_lambda}), then the increase in the number of devices does not provide any gain for the system performance, but with an increase in the EH rate, the gain increases as the system size goes from $n = 1$ to $n = 2$ if $\lambda \leq 0.6$. If $\lambda \geq 0.4$ the gain obtained by an increase of the system size from $n = 1$ to $n = 2$ and from $n = 2$ to $n = 4$ is similar. Nevertheless, in the performance comparison in case of $n = 1$ we consider the best performing setting, i.e., $c_1 = 1$, $p_1 = 0.1$. If $c_1 >1$, $p_1 >0.1$, the gain is much more significant when the system size is increased from $n = 1$ to $n = 2$.
A similar statement holds true when we vary the values of energy arrivals, $\bar{e}$. This is because, when the EH rate is low, the monitoring node almost exclusively uses the ``cheaper'' information sources, so introducing more expensive alternatives does not help.

\begin{figure}[!t]	
	\centering
	\includegraphics[width=0.92\columnwidth]{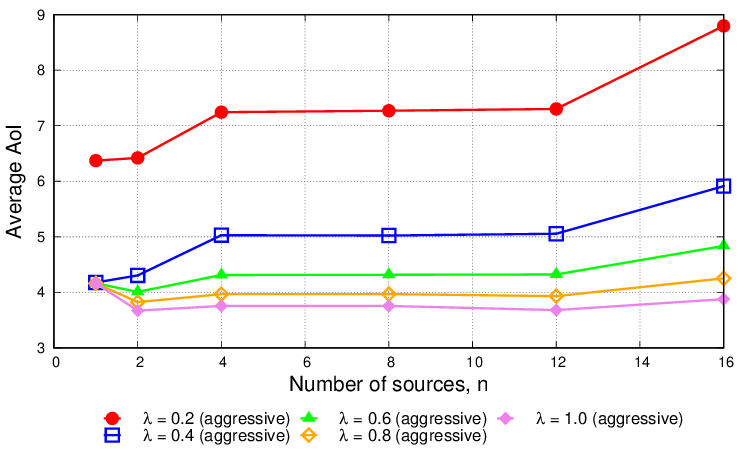}
	\caption{Average AoI vs. number of information sources vs.  if the monitoring node adopts aggressive strategy for different values of the EH rate, $\lambda$.}
	\label{fig:number_aggressiveAoI_lambda}	

\end{figure}

If the monitoring node exploits the aggressive strategy, then we observe a counterintuitive behaviour: with an increase in the system size for $n > 2$, the performance worsens, or, in other words, the average AoI at the monitoring node increases. Moreover, the lower $\lambda$, the higher the increase in the average AoI, or the more inefficient the aggressive policy becomes. This effect is particularly negative for low $\lambda$ values because the aggressive policy always goes for the most expensive information source it can afford. Introducing more expensive alternatives means that they will end up being used rather  than the cheaper sources. This results in a poorer performance particularly for low $\lambda$ values, when it is optimal to exploit only the cheapest sources. Yet, when $\lambda \geq 0.4$ introducing system diversity slightly improves the performance; actually, the best performance is provided by $c_1 = 8$  (see Fig.\ \ref{fig:one_source}), therefore, when we shift from $C=[1]$ to $C = [16, 1]$ we achieve the ``balance'' and an improvement in the performance. However, with  a further increase of the system size the ``balance'' shifts causing an increase in average AoI (see Fig. \ref{fig:number_aggressiveAoI_lambda}).

\section{Conclusions}
\label{sec:con}

In this work, we considered a system model with a single energy-harvesting monitoring node that can request status updates from multiple heterogeneous information sources that monitor the same process of interest. We assumed that the energy cost of requesting an update, as well as the statistics of the age of the received update varies across the information sources. In order to analyze the system, we considered different combinations of costs and age distributions that are described in detail in Section \ref{sec:res}.

We formulated the long-term average AoI minimization problem as an MDP, and obtained the optimal solution using the relative VI algorithm. We then studied the optimal solution for different EH rates and found out that the solutions are more sensitive to the age distribution rather than the costs of the status updates. We demonstrated that having just the cheapest sources is mostly sufficient if the EH rate is low. We also considered an aggressive policy, which requests a status update from the most expensive source it can afford at each time slot, as a benchmark. We observed that the aggressive policy is near optimal when the EH rate is high. 

We found that adding information sources beyond a certain number does not help, particularly if the available sources already provide sufficient diversity in terms of the cost-average age trade-off within the available energy sources. 

Future work includes an extended model comprising the channel dynamics
 and the resulting transmission time and costs, as well as more general EH schemes \cite{priya2009energy}. 
Another direction is to study reinforcement learning to choose the information source to use over time without depending on the explicit information on the age distributions of the sources or the statistics of the EH process \cite{abd2020reinforcement}.
\appendices
\section{Proof of Lemma \ref{lemma:2}}
\label{sec:proof_1}
The lemma requires to prove, for any $b$, the existence of a $\vartheta_1(b)$ such that it is convenient to update if and only if $\delta \geq \vartheta_1(b)$.
First of all, for $b < b_0 = \min_{j = 1,2,\dots,N} c_j$, then the statement is trivially true with an infinite $\vartheta_1(b)$ as all sources are too expensive to update.

Define $\bar{R}(s(t) {=} (b, \delta),a_i)$  as the average optimal reward starting from the current state $s(t){=}(b,\delta)$ and after taking action $a(t) {=} a_i$. According to our model, if $i>0$, that is, we perform an actual update from source $i$, we evolve to a state with either energy level $b{-}c_i$ or $b{-}c_i{+}\bar{e}$ (depending on the EH process) and AoI $\epsilon {\in} \mathcal{E}_i {=} [\alpha, \min(\beta, \delta{+}1)
]$ with probability $\tilde{\gamma}_i(\epsilon)$, which is defined as
follows. If $\beta {\leq} \delta$ then $\tilde{\gamma}_i(\epsilon) =
\gamma_i(\epsilon)$ for all $\epsilon$. If $\beta {>} \delta$ then
$\tilde{\gamma}_i(\epsilon) = \gamma_i(\epsilon)$ for $\epsilon \in
[\alpha,\delta]$ and $\tilde{\gamma}_i(\delta{+}1) = \sum_{n =
	\delta+1}^{\beta} \gamma_i(n)$.

From (\ref{tr_a0}) and (\ref{eq:reward}) we can write the following Bellman equation for $\bar{R}(s(t),a_i)$ being
\allowdisplaybreaks
\small
\begin{multline}
		\max_{j \in  \Omega}  \bigg[ (1{-}\lambda) \sum_{\epsilon \in \mathcal{E}_i } \!\! \Big(\epsilon {+} \tilde{\gamma}_i(\epsilon) \bar{R}\big(s(t{+}1){=} (b{-}c_i,\epsilon), a_j\big) \! \Big)  
		+\lambda \sum_{\epsilon \in \mathcal{E}_i } \!\! \Big(\epsilon {+} \\ \tilde{\gamma}_i(\epsilon)  \bar{R}\big(s(t{+}1){=} (b{-}c_i+\bar{e},\epsilon), a_j\big)\! \Big)\bigg],  \text{\small where } \Omega {=} \{0, \dots, n\}
		\label{eq:leo0}
		\vspace{-0.5cm}
\end{multline}
\normalsize
Given that $\epsilon$ does not depend on $a_j$,  (\ref{eq:leo0}) can be written as: 
\allowdisplaybreaks

\begin{multline}
	(1{-}\lambda) \sum_{\epsilon \in \mathcal{E}_i } \!\! \Big(\epsilon {+} \tilde{\gamma}_i(\epsilon) \max_{j \in \Omega} \bar{R}\big(s(t{+}1){=} (b{-}c_i,\epsilon), a_j\big) \! \Big) 
	\\
	+\lambda \sum_{\epsilon \in \mathcal{E}_i } \!\! \Big(\epsilon {+} \tilde{\gamma}_i(\epsilon) \max_{j \in  \Omega} \bar{R}\big(s(t{+}1){=} (b{-}c_i+\bar{e},\epsilon), a_j\big)\! \Big)
	\label{eq:leo1}
\end{multline}

whereas if we do not update (action $a_0$) we obtain
\begin{multline}
	\small
	\bar{R}(s(t),a_0) {=} (1{-}\lambda) \Big( \delta  {+}  \max_{j \in  \Omega}\bar{R}\big(s(t{+}1){=}(b,\delta{+}1), a_j\big) \! \Big)  \\
	 +\lambda \Big( \delta  {+}  \max_{j \in  \Omega} \bar{R}\big(s(t{+}1){=} (b{+}\bar{e},\delta{+}1), a_j\big) \! \Big)
	\label{eq:leo2}
\end{multline}

We notice that (\ref{eq:leo1}) is not explicitly influenced by $\delta$, which is incidentally logical as, after the update, $\delta$ is reset to a ``low'' AoI value,\footnote{To be precise, the set $\mathcal{E}_i$ actually depends on $\delta$ but only for the reason that whenever the update is supposed to be to a value in $[\alpha, \beta ]$ that is higher than $\delta+1$, the update information is actually useless and discarded, see (\ref{eq:updatedelta}).} whereas (\ref{eq:leo2}) is increasing in $\delta$.
This implies that as $\delta$ increases,  there exists a turning point $\vartheta_1(b)$ that makes (\ref{eq:leo1}) smaller
than (\ref{eq:leo2}) and this stays true for all values of $\delta \geq
\vartheta_1(b)$. \color{black}

\section{Proof of Lemma \ref{theorem_2}}
\label{sec:proof_2}
\vspace{-0.1cm}
Similarly to the previous lemma, we can compare the Bellman equations
for the updates from two different sources $i$ and $j$. Note that we
are always considering AoI values $\delta > \vartheta_1(b)$ for which
an update is preferable to staying idle, as proven in Lemma 1. And
since we are updating in both cases, we lose any explicit dependence
on the AoI $\delta$, as per (\ref{eq:leo1}) - in other words, after either update
action, the system trajectory evolves from states with ``low'' AoI.
Finally, we remark that $\bar{R}\big( (b,\delta), a \big)$ is always
non-decreasing in $\delta$ for given $b$ and action $a$.
This implies that if $\bar{R}\big( (b,\delta_1), a_i \big) <
\bar{R}\big( (b,\delta_1), a_j \big)$ but
$\bar{R}\big( (b,\delta_2), a_i \big) > \bar{R}\big( (b,\delta_2), a_j
\big)$, then
$\bar{R}\big( (b,\delta), a_i \big) > \bar{R}\big( (b,\delta), a_j
\big)$ also for every $\delta>\delta_2$; that is, a source to
update from can be the optimal one only over a set of contiguous AoI values.


\vspace{-0.3cm}
\bibliographystyle{IEEEtran}

%
\vspace{-.5cm}
\begin{IEEEbiography}[{\includegraphics[width=\columnwidth]{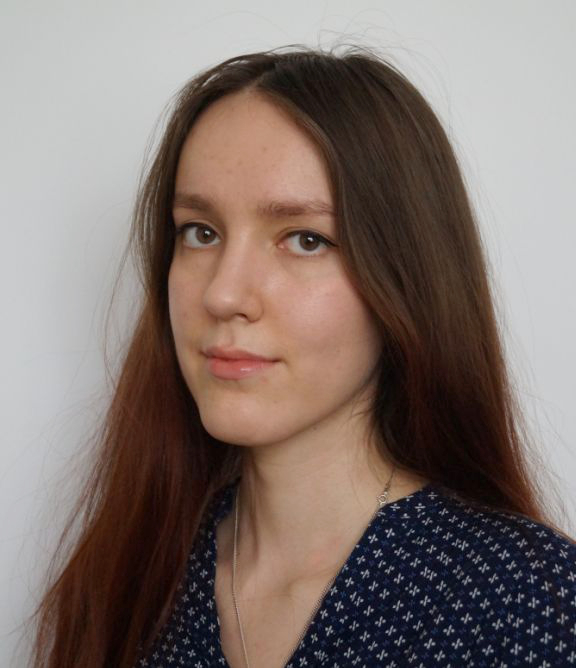}}]{Elvina Gindullina}
	received her M.S. degree (with honor) in applied mathematics and computer science from Ufa State Aviation technical University (USATU), Russia, in 2015. From 2014 to 2016 she joined USATU as a programmer.  In 2016, she moved to Italy where she received the
	PhD degree in Information and communication technologies  from University of Padova (UNIPD), Italy, in 2020; In  2016 she joined SCAVENGE project (Horizon 2020) as an early stage researcher (ESR), under the Marie Skłodowska-Curie Actions programme, where she was working on battery management systems, energy cooperation solutions and increasing of energy efficiency for energy harvesting IoT networks. During  2018 she conducted her internship in Worldsensing (Barcelona, Spain) as a research engineer, where she was designing sampling strategies for an industrial data-logger powered by a solar panel. She is currently a research assistant in UNIPD  (Italy) in the Department of Information Engineering  where she is developing and analysing the novel models and methods for effective connectivity in a whole-brain network.
\end{IEEEbiography}
\vskip 0pt plus -1fil
\begin{IEEEbiography}[{\includegraphics[width=\columnwidth]{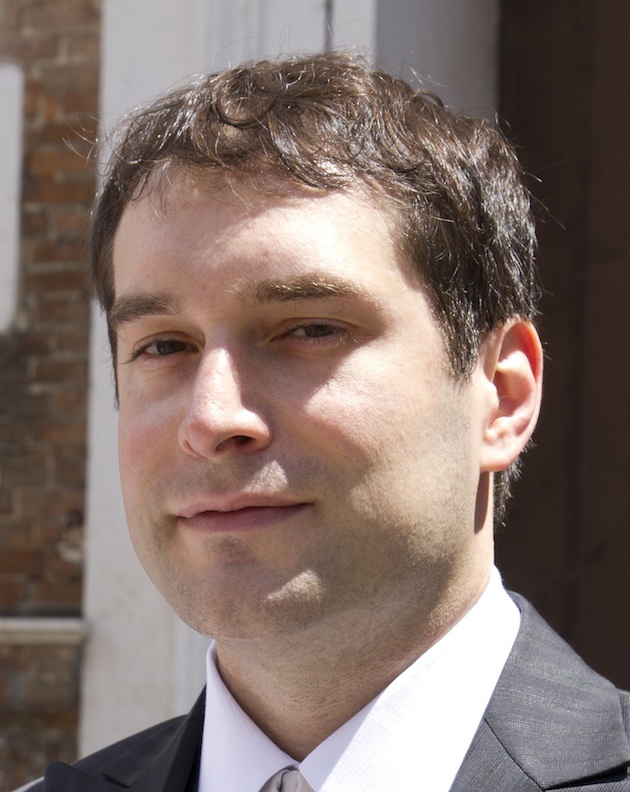}}]{Leonardo Badia}
	Leonardo Badia received the Laurea degree (with honors) in electrical engineering and the PhD degree in information engineering from the University of Ferrara, Italy, in 2000 and 2004, respectively. 
	During 2002 and 2003, he was on leave at the Radio System Technology Labs (now Wireless@KTH), Royal Institute of Technology of Stockholm, Sweden. After having been with the Engineering Department of the University of Ferrara, he joined in 2006 the IMT Institute for Advanced Studies, in Lucca, Italy. In 2011, he moved to the University of Padova, Italy, where he is currently an Associate Professor. His research interests include protocol design for multi-hop networks, cross-layer optimization of wireless communication, transmission protocol modeling, and applications of game theory to radio resource management. 
	Professor Badia published more than 140 research papers. He is involved in the coordination of scientific events and projects. He is also an active referee of research articles, having served on the editorial boards and still being an active reviewer for many scientific periodicals, as well as technical program committee chair for conferences in the broad area of networking and wireless communications.
	
\end{IEEEbiography}
\vskip 0pt plus -1fil
\begin{IEEEbiography}[{\includegraphics[width=\columnwidth]{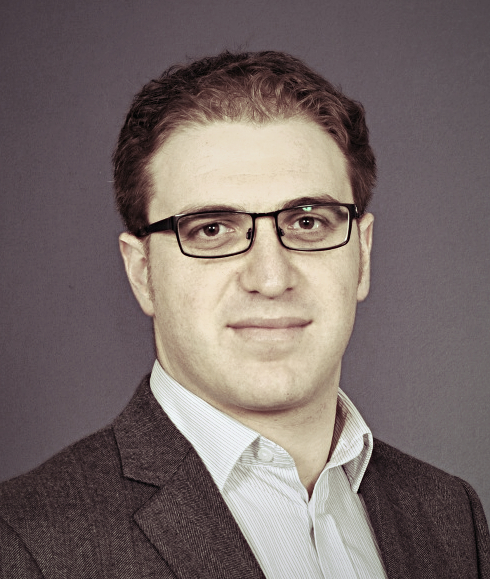}}]{Deniz G\"{u}nd\"{u}z}
	received M.S. and Ph.D. degrees in electrical engineering from NYU Tandon School of Engineering (formerly Polytechnic University) in 2004 and 2007, respectively. After his PhD, he served as a postdoctoral research associate at Princeton University, and as a consulting assistant professor at Stanford University. From Sep. 2009 until Sep. 2012 he served as a research associate at CTTC in Barcelona, Spain. ln Sep. 2012, he joined the Electrical and Electronic Engineering Department of Imperial College London, UK, where he is currently a Professor in Information Processing, serves as the deputy head of the Intelligent Systems and Networks Group, and leads the Information Processing and Communications Laboratory (IPC-Lab). He is also a part-time faculty member at the University of Modena and Reggio Emilia, and has held visiting positions at University of Padova (2018-2020) and Princeton University (2009-2012).
	His research interests lie in the areas of communications and information theory, machine learning, and privacy. Dr. Gunduz is an Area Editor for the IEEE Transactions on Communications and the IEEE Journal on Selected Areas in Communications (JSAC) - Special Series on Machine Learning in Communications and Networks. He also serves as an Editor of the IEEE Transactions on Wireless Communications. 
	%
\end{IEEEbiography}

\end{document}